\documentclass[acmlarge]{acmart}

\usepackage{subfigure}
\graphicspath{{image/}}

\usepackage{scalerel}
\newcommand {\inlinefig}[1] {\scalerel*{\includegraphics{#1}}{|}}

\usepackage{algorithm2e}

\AtBeginDocument{%
  \providecommand\BibTeX{{%
    \normalfont B\kern-0.5em{\scshape i\kern-0.25em b}\kern-0.8em\TeX}}}

\setcopyright{acmcopyright}
\copyrightyear{2020}
\acmYear{2020}
\acmDOI{10.1145/1122445.1122456}

\acmJournal{TOCS}
\acmVolume{36}
\acmNumber{4}
\acmArticle{111}
\acmMonth{6}

\begin{document}

\title[CYBEX-P \& TAHOE]{Cybersecurity Information Exchange with Privacy (CYBEX-P) and TAHOE -- A Cyberthreat Language}

\author{Farhan Sadique}
\email{fsadique@nevada.unr.edu}
\orcid{0000-0002-2412-1857}
\author{Ignacio Astaburuaga}
\email{ignacio.ag@nevada.unr.edu}
\author{Raghav Kaul}
\email{raghavkaul@nevada.unr.edu}
\author{Shamik Sengupta}
\email{ssengupta@unr.edu}
\author{Shahriar Badsha}
\email{sbadsha@unr.edu}
\author{James Schnebly}
\email{jschnebly@nevada.unr.edu}
\author{Adam Cassell}
\email{acassell@nevada.unr.edu}
\author{Jeff Springer}
\email{jeffs@unr.edu}
\author{Nancy LaTourrette}
\email{nancy@unr.edu}
\author{Sergiu M. Dascalu}
\email{dascalus@cse.unr.edu}
\affiliation{
  \institution{University of Nevada, Reno}
  \streetaddress{1664 North Virginia Street}
  \city{Reno}
  \state{Nevada}
  \postcode{89557}
}

\renewcommand{\shortauthors}{Sadique, et al.}

\begin{abstract}
  Cybersecurity information sharing (CIS) is envisioned to protect organizations more effectively from advanced cyberattacks. However, a completely automated CIS platform is not widely adopted. The major challenges are: (1) the absence of a robust cyberthreat language (CTL) and (2) the concerns over data privacy. This work introduces Cybersecurity Information Exchange with Privacy (CYBEX-P), as a CIS framework, to tackle these challenges. CYBEX-P allows organizations to share heterogeneous data with granular, attribute based privacy control. It correlates the data to automatically generate intuitive reports and defensive rules. To achieve such versatility, we have developed TAHOE -- a graph based CTL. TAHOE is a structure for storing, sharing and analyzing threat data. It also intrinsically correlates the data. We have further developed a universal Threat Data Query Language (TDQL). In this paper, we propose the system architecture for CYBEX-P. We then discuss its scalability and privacy features along with a use case of CYBEX-P providing Infrastructure as a Service (IaaS). We further introduce TAHOE \& TDQL as better alternatives to existing CTLs and formulate ThreatRank -- an algorithm to detect new malicious events.
\end{abstract}


\maketitle

\section{Introduction}\label{sec:intro}

Collaborative cybersecurity information sharing (CIS) is envisioned to protect organizations more effectively from advanced cyberattacks \cite{feledi2013toward}. The benefits of proactive sharing are twofold --- (1) new threats are detected faster, owing to collaborative analysis (2) the corresponding signatures are distributed faster due to real-time sharing.

Moreover, it has been repeatedly showed using game theoretic modeling that organizations are benefited by sharing threat data (e.g.  malware signatures or firewall logs) \cite{tosh2015evolutionary, vakilinia2017privacy}. Furthermore, the US congress proposed a number of acts incentivizing private organizations and requiring public organizations to share threat data \cite{cisa2014, cistc2014}.

\subsection{Motivation and Challenges}

Despite all the benefits, there is limited sharing in the industry due to several limitations of existing platforms:

\begin{enumerate}
	\item The existing platforms are built for only data sharing \cite{connolly2014trusted} or for limited data analysis \cite{wagner2016misp}, although robust data analysis is just as important \cite{vazquez2012conceptual},
	\item They cannot generate actionable CTI from machine data \cite{wagner2016misp}; automatic data collection is either absent \cite{dandurand2013towards} or limited \cite{sauerwein2017threat,sillaber2016data} in existing platforms, although complete automation is expected \cite{vazquez2012conceptual,kijewski2014proactive},
	\item There is no standard cyberthreat language for all of data sharing, storing, correlation, and analysis \cite{vazquez2012conceptual}
	\item There are no intuitive investigation tools \cite{vazquez2012conceptual}, and
	\item The existing platforms cannot outline a defensive course of action - e.g. automatic generation of firewall rules, although it is desired in a complete CIS ecosystem \cite{mcconnell2011and}.
\end{enumerate}

The situation is further aggravated by several risks associated with sharing private information because:

\begin{enumerate}
	\item it may reveal vulnerabilities in the sharer's network attracting more targeted attacks \cite{chismon2015threat},
	\item it can compromise the privacy of the users \cite{sharma2014critter},
	\item it may violate existing data policy of organizations \cite{sander2015ux},
	\item it potentially subjects organizations to government surveillance \cite{burger2014taxonomy},
	\item revealing vulnerabilities may damage an organization's reputation \cite{garrido2016shall}, and
	\item competitors may acquire significant underlying intelligence from the data \cite{chismon2015threat}.
\end{enumerate}

To tackle these challenges, we introduce CYBersecurity information EXchange with Privacy (CYBEX-P) in this paper. CYBEX-P is a cybersecurity information sharing (CIS) platform with robust data governance. It automatically analyzes shared data to generate insightful reports and alerts.

We further introduce TAHOE -- a cyberthreat language (CTL), which correlates new events with older ones to predict future attacks. Moreover, CYBEX-P provides infrastructure as a service (IaaS) for threat data analysis. Our experimental results show that CYBEX-P is scalable and suitable for real-time networks. Thus, CYBEX-P can disrupt the rapid and extensive spread of new threats.

\subsection{Contribution}

The novel contributions of this work are:

\begin{enumerate}
	\item CYBEX-P -- a completely automated CIS framework for with data collection, data analysis, privacy preservation and report generation,
	\item TAHOE -- a cyberthreat language (CTL) for storing, sharing, analyzing and intrinsically correlating data,
	\item TDQL -- a universal Threat Data Query Language to query any threat data from any database,
	\item ThreatRank -- a novel algorithm to detect previously unseen malicious events using correlation,
	\item A privacy preservation mechanism that provides granular, attribute-based access control. It also allows CYBEX-P to correlate the encrypted data without exposing them,
	\item A report/alert module to automatically generate preventive rules (e.g. firewall rules),
	\item CYBEX-P Threat Intelligence -- a graphical tool to visualize and investigate incidents, and,
	\item A real-time phishing URL detector that uses CYBEX-P infrastructure as a service (IaaS).
\end{enumerate}

\section{Related Work}\label{sec:rw}

While there are plenty of works on cybersecurity information sharing (CIS), none of them provide a comprehensive solution to the aforementioned challenges. In this section we discuss these works, focusing on the CIS frameworks.

We begin our study with CIS frameworks proposed in academia. Edwards et al. \cite{edwards2002system} presented one of the earliest frameworks for sharing vulnerability information. Another framework was presented by Zhao et al. \cite{zhao2012collaborative} for collaborative information sharing. Yet another framework, called SKALD, \cite{webster2016skald} was developed by Webster et al., for real-time sharing. However, none of these preserve the privacy of the data making them undesirable \cite{chismon2015threat, sharma2014critter, sander2015ux, burger2014taxonomy, garrido2016shall, chismon2015threat}.  CYBEX-P sets itself apart from these early works by providing a robust system architecture along with a novel privacy preservation mechanism.

Meanwhile, several proprietary frameworks have emerged in the industry including ThreatConnect \cite{platforms2017threatconnect}, AlienVault \cite{ossim25open}, X-Force \cite{ibmxforce}, ThreatStream \cite{anomalithreatstream}, ThreatExchange \cite{fbthreatexchange}, and EclecticIQ \cite{eclecticiq}. All of these suffer from one or more of three major limitations: (1) the data are inputted by human not automated (2) privacy of shared data is not preserved (3) they have limited scope in participant or type of data. For example, ThreatExchange does not allow educational institutions, X-Force data are written by humans and so on. Incidentally, none of these allow encrypting the shared data for privacy preservation. CYBEX-P, on the other hand, is built from the ground up keeping privacy in mind.

Now, we examine the frameworks which do consider privacy. Goodwin et al. \cite{goodwin2015framework} were one of the first to coin the features of a comprehensive CIS framework. In their work, they recommended developing an extensive framework with privacy preservation and data governance. Although, they outlined really well what needs to be done, they did not clarify how to do it. Furthermore, our work does not rely on voluntary data sharing as proposed in \cite{goodwin2015framework}. We incentivize data sharing by mitigating risk (by providing analysis reports).

Another, privacy preserving framework called PRACIS \cite{de2017pracis} was introduced by de Fuentes et al. They generated several summary statistics by aggregating homomorphically encrypted data. However, they did not propose a comprehensive system architecture with heterogeneous data collection and automated data analysis. 

Now, we move onto the most prominent CIS systems in use today. The primary requisite of any CIS platform is a standardized format or a cyberthreat language (CTL). Presently, the most popular CTL is Structured Threat Information Expression (STIX) \cite{barnum2012standardizing} developed by the MITRE corporation. While STIX is perfect for mutual data sharing, it is unscalable for any kind of data analysis. Moreover, \ref{sss:dup} discusses how STIX based systems are prone to store duplicate data. To overcome these shortcomings, \ref{sec:tahoe} introduces TAHOE - a graph based CTL for both data sharing and data analysis.

In parallel to STIX, MITRE also developed the Trusted Automated eXchange of Indicator Information (TAXII) \cite{connolly2014trusted} protocol to facilitate peer-to-peer data sharing between trusted parties. However, what TAXII gains in data sharing, it lacks in privacy. For example, it does not provide attribute-based access control. In contrast, CYBEX-P is built from a privacy standpoint giving data owners complete control to choose who sees which specific attribute. Furthermore, since TAXII only supports STIX format, it is unsuitable for even the simplest data analysis. For example, TAXII does not provide any API endpoint to lookup an IP address in its database..

Meanwhile, Wagner et al. developed MISP \cite{wagner2016misp} as a collaborative CTI sharing platform with group based access control and their own data format. Although, database lookups are fast in a MISP server, their open data structure defeats that purpose, because different users structure the same data in different ways. Moreover, MISP structures data in only two levels (Events and Attributes); so representing complex data in MISP format is non-intuitive. A few other limitations of MISP are: it relies heavily on manual human input rather than automating machine data, and it does not provide a robust data governance framework.

In summary, after extensive study, we were primarily inspired by both STIX and MISP data structures while developing TAHOE. However, we have built TAHOE from the ground up with privacy and speed in mind. Similarly, we were inspired by both TAXII and MISP while designing CYBEX-P. However, CYBEX-P is a complete ecosystem with actionable data rather than a cumbersome tool.  As a result, we present TAHOE and CYBEX-P as a perfect marriage between versatility and performance in this paper.

\section{Overview of CYBEX-P}\label{sec:cybexp}

\begin{figure}[h]
    \centering
	\includegraphics[height=2in]{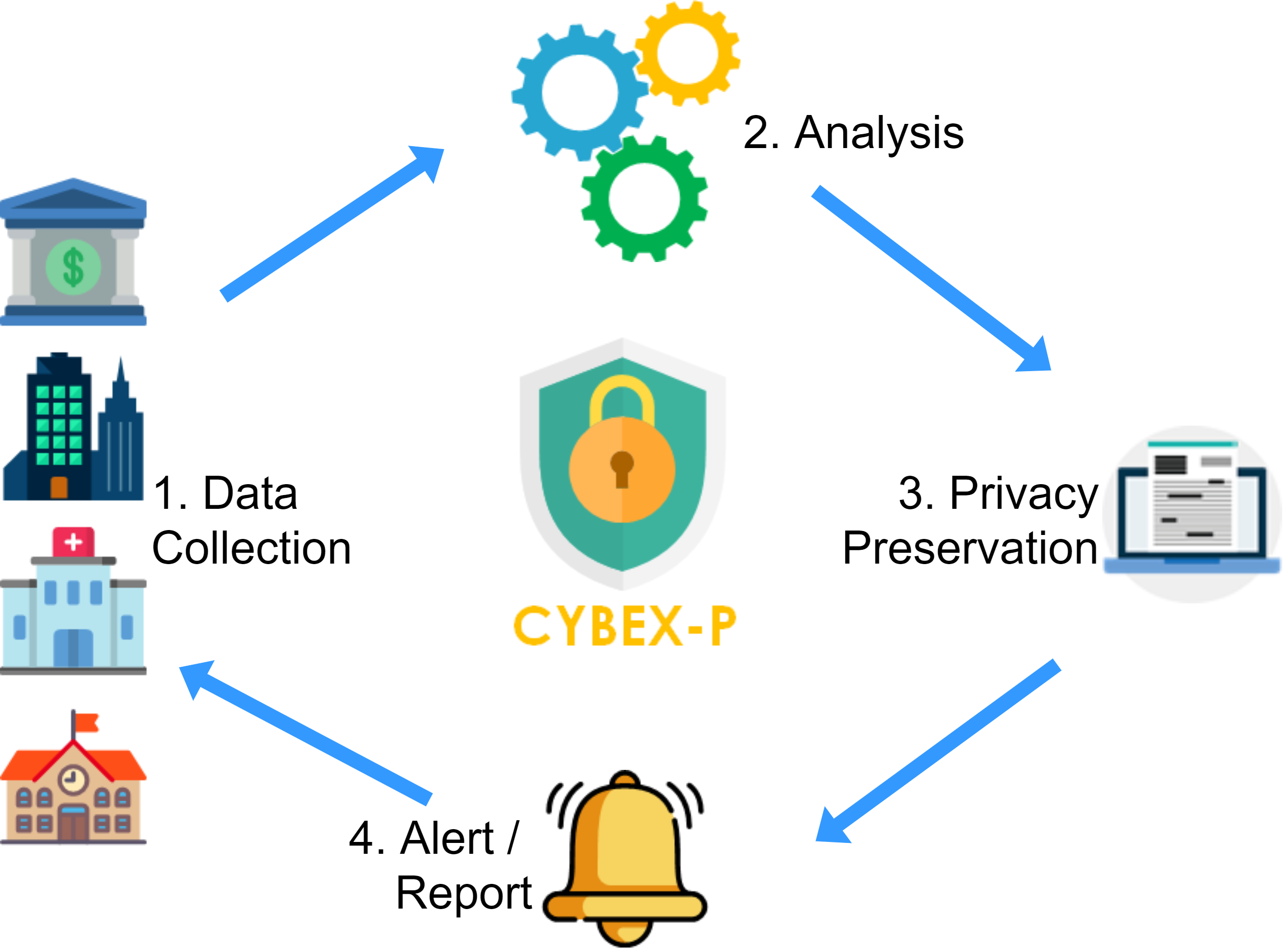}
	\caption{Overview of the $4$ major functions of CYBEX-P.}
	\label{fig:overview}
    \Description{ss.}
\end{figure}

We begin our discussion with a functional overview of CYBEX-P. Fig. \ref{fig:overview} shows CYBEX-P's $4$ major functions.

\subsection{Data Collection}
CYBEX-P is essentially a cloud based platform for organizations to share heterogeneous cyberthreat data. CYBEX-P accepts all kinds of human or machine generated data including firewall logs, emails, malware signatures, and handwritten cyberthreat intelligence (CTI).

\subsection{Data Analysis}
In addition to data sharing, CYBEX-P allows the users to correlate and analyze the data. This key feature sets CYBEX-P apart from other cybersecurity information sharing systems.

\subsection{Privacy Preservation}
The second key feature of CYBEX-P is that, the data owner controls, who sees which part of the data. We achieve such a granular control by separately encrypting each attribute of the data. The privacy preservation mechanism is illustrated in section \ref{sec:datagov}.

\subsection{Report/Alert Generation}
Finally, users can generate insightful reports or alerts from the data. CYBEX-P also provides a feed of automatically generated defensive (e.g. firewall) rules, as we will discuss in \ref{sss:rule}. This particular feature reflects our philosophy of making the entire process completely automated.


\section{System Architecture of CYBEX-P}\label{sysarch}

\begin{figure}[h!]
	\includegraphics[width=.6\linewidth]{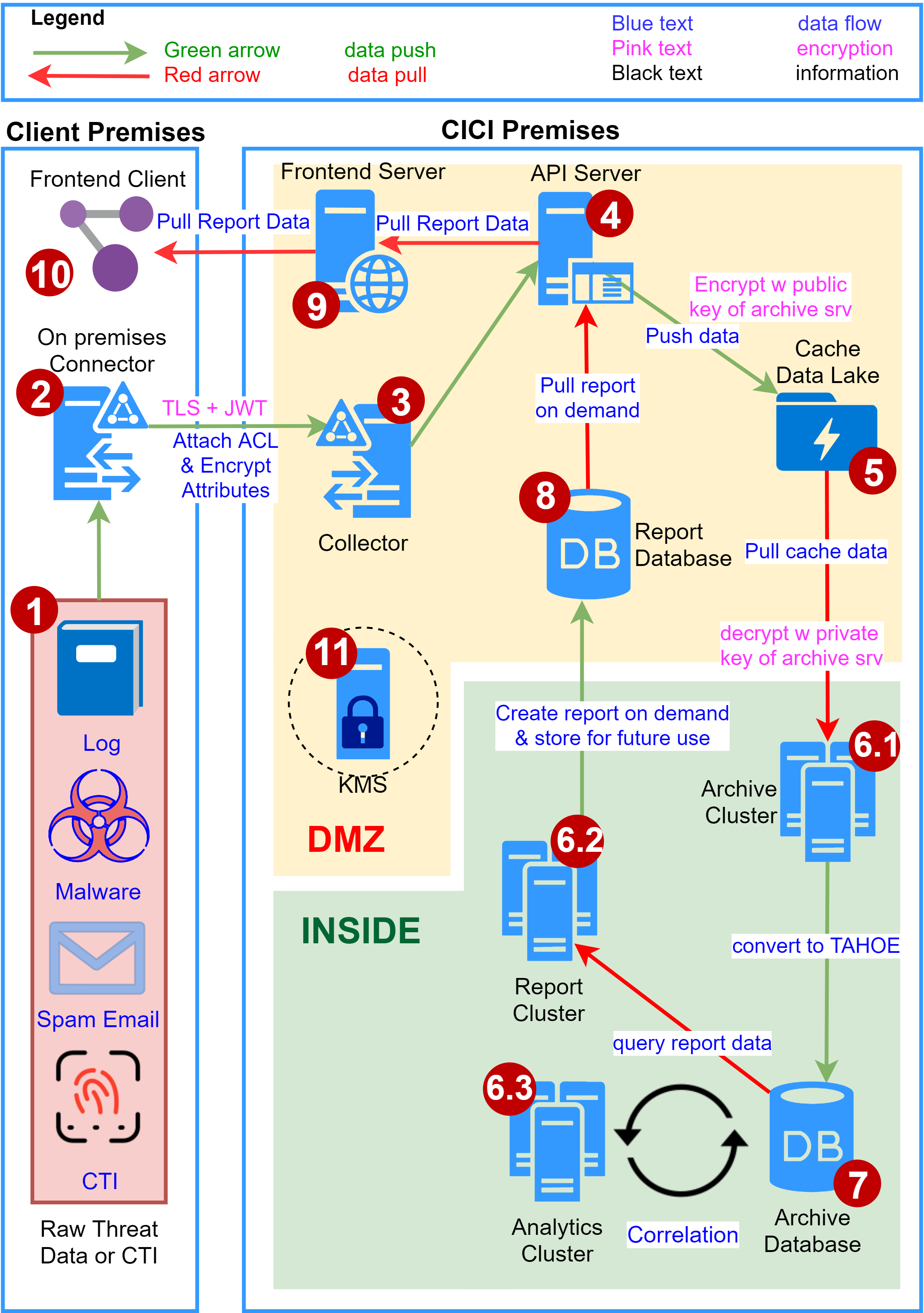}
	\centering
	\caption{System architecture of CYBEX-P along with the Data Flow.}
	\label{fig:sysarch}
\end{figure}

To accommodate the four major functions, we have built CYBEX-P with $6$ independent software modules -- (1) Frontend, (2) Input, (3) API, (4) Archive, (5) Analytics, and (6) Report. These modules share various components as shown in Fig. \ref{fig:sysarch}. In addition, we have built a library to manipulate TAHOE content. TAHOE is a CTL, that CYBEX-P uses, to store, analyze and share data.


\subsection{Frontend Module}\label{ss:fend}

The frontend module (\inlinefig{9}, \inlinefig{10} in Fig. \ref{fig:sysarch}) is a webapp for users to interact with CYBEX-P. This module allows users --

\begin{enumerate}
	\item to register with and login to CYBEX-P,
	\item to manually upload threat data as text files,
	\item to configure machines (e.g. firewalls) to automatically share data with CYBEX-P (explained in \ref{subsec:input}),
	\item to control the access of their data (explained in \ref{sec:datagov}),
	\item to generate and view reports (explained in \ref{subsec:report}), and
	\item to investigate an incident using our incident investigation tool (explained below).
\end{enumerate}

\subsubsection{CYBEX-P Threat Intelligence -- An Incident Investigation Tool}\label{sss:neo4j}

\begin{figure}[h]
	\includegraphics[width=\linewidth]{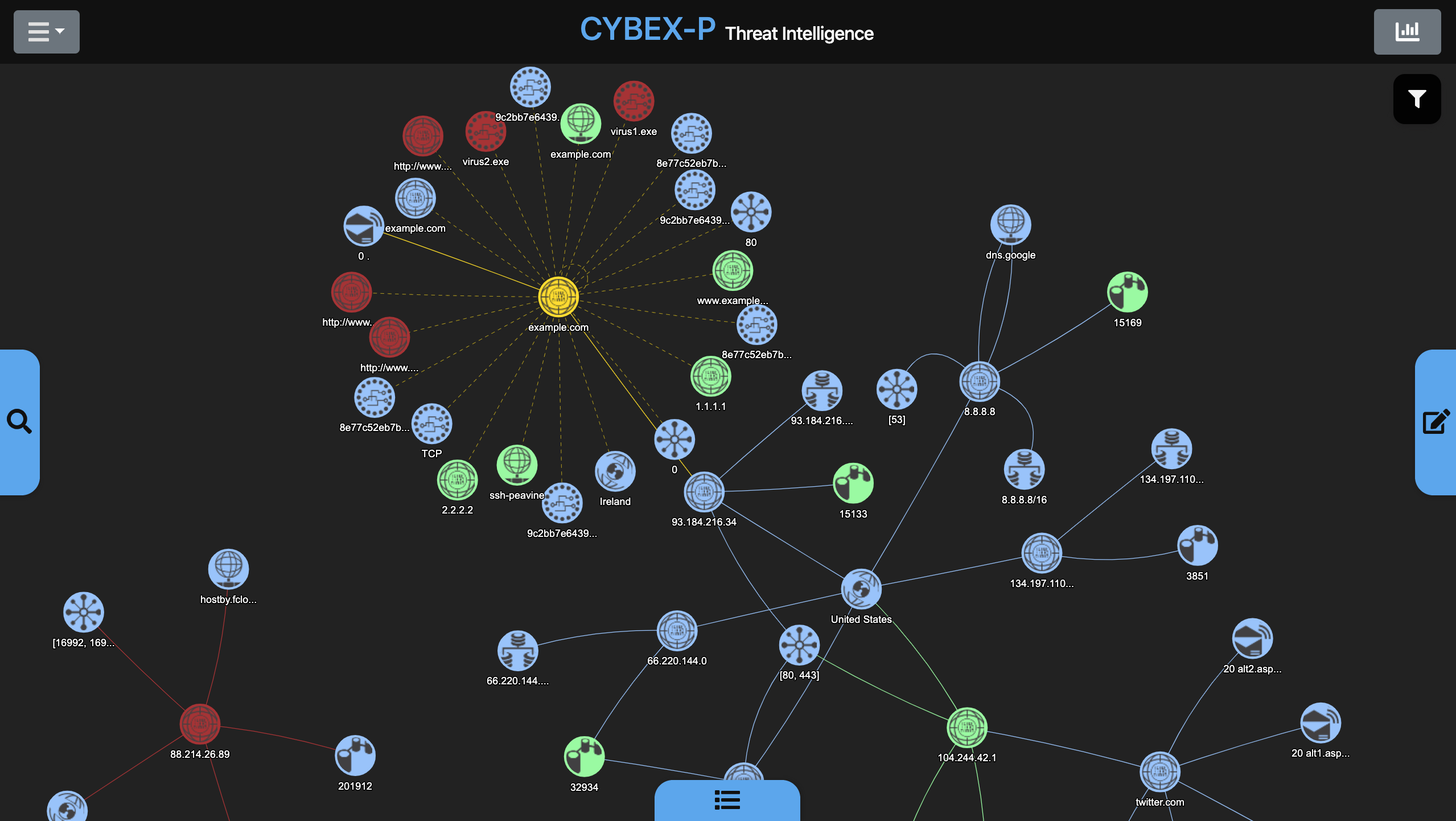}
	\centering
	\caption{Graphical Incident Investigation Tool.}
	\label{fig:investigation}
\end{figure}

Fig. \ref{fig:investigation} shows a novel contribution of this project -- CYBEX-P Threat Intelligence. This tool, powered by CYBEX-P analytics, allows a user to investigate an incident.

The investigation starts with a blank canvas. Firstly, the user inputs one or more attributes to the canvas. The attributes become nodes or vertices in the graph.

Secondly, the user clicks a button to enrich the graph with related attributes. This tool uses CYBEX-P database to get the related attributes. Related attributes are discussed in \ref{sec:tahoe}.

Thirdly, CYBEX-P sends a malicious score for all the attributes on the graph. Section \ref{sec:pagerank} explains ThreatRank -- the algorithm used to calculate these malicious scores.

Finally, the investigation tool colors each attribute blue (unknown), green (benign), yellow (suspicious) or red (malicious) based on the score. A big cluster of red attributes denotes that the original attribute is malicious.

\subsection{Input module}\label{subsec:input}

The input module (\inlinefig{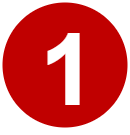}, \inlinefig{2}, \inlinefig{3}, \inlinefig{4}, \inlinefig{10} in Fig. \ref{fig:sysarch}) handles all kinds of data incoming to CYBEX-P. Users can manually upload threat data via a web client (\scalerel*{\includegraphics{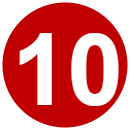}}{|}) or automatically send machine data via a connector (\inlinefig{2}) to the collector (\scalerel*{\includegraphics{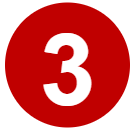}}{|}).

The collector automatically collects heterogeneous data from any source e.g. a firewall, an email forwarder etc. Example methods for automatic data collection are -- (1) by calling an API, (2) via a pre-configured websocket, (3) by reading from a text file, (4) by reading from a database, (5) using Linux syslog protocol etc.



Afterwards, the collector posts the raw data to our API (\scalerel*{\includegraphics{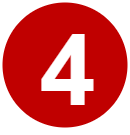}}{|}) endpoint. To ensure privacy, it uses the transport layer security (TLS) protocol \cite{dierks2008transport} during collection and posting.

\subsection{API module}\label{subsec:api}

The API module (\inlinefig{4}, \inlinefig{5} in Fig. \ref{fig:sysarch}) consists of the API server (\inlinefig{4}) and the cache data lake (\inlinefig{5}). It acts as the gateway for all data into and out of CYBEX-P. It has two sub-modules --



\subsubsection{Data Input sub-module}

The input module posts the raw data to the API (\scalerel*{\includegraphics{4}}{|}) endpoint. The API encrypts the data with the public key of the archive server (\inlinefig{6.1}) and stores the encrypted data in the cache data lake (\scalerel*{\includegraphics{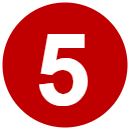}}{|}).

We have placed the API in the demilitarized zone (DMZ) of our firewall, because it faces the internet. However, storing data in the DMZ is somewhat risky. So, we encrypt the cache data lake with the public key of the archive server. The archive server is in the inside zone. This design protects that data even if the DMZ is compromised.



\subsubsection{Report Publishing sub-module}

A user can request different reports via the API. The API gets those reports from the report DB (\scalerel*{\includegraphics{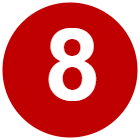}}{|}) and presents them to the user. Thus, the API module acts as an interface for all data.

\subsection{Archive module}\label{subsec:archive}

The archive module (\inlinefig{6.1}, \inlinefig{7} in Fig. \ref{fig:sysarch}) resides in the archive cluster and consists primarily of a set of parsing scripts. As mentioned earlier, the cache data lake (\inlinefig{5}) is encrypted with the public key of the archive server (\inlinefig{6.1}). The archive server -- (1) gets the encrypted data from the cache data lake (2) decrypts the data using own private key (3) parses the data into TAHOE, and (4) stores the data in the archive DB (\inlinefig{7}).

\subsubsection{Performance Challenge}

The archive module potentially handles hundreds of different data formats from thousands of sources. It is reconfigured every time CYBEX-P connects to a new data source. Moreover, it checks each piece of new data against the entire database to determine if it's a duplicate (explained in subsubsection \ref{sss:dup}).




\subsubsection{Design Choices}

To tackle these challenges, we have made the archive module separate so that we can reconfigure it without affecting other modules. We have also placed it between two databases (cache data lake and archive DB) so that we don't lose any data when we reconfigure it. Finally, we have employed parallel computing, because our data schema (TAHOE) considers each piece of data independent of another.





\subsection{Analytics module}\label{subsec:analytics}

The analytics module (\inlinefig{6.3}, \inlinefig{7} in Fig. \ref{fig:sysarch}) works on the archived data to transform, enrich, analyze or correlate them. It has various sub modules, some of which described here.

\subsubsection{Filter sub-module}

An analytics filter parses a specific event from raw user data. Multiple filters can act on the same raw data and vice-versa.  For example, one filter can extract a \textit{file download event} from a piece of data while another filter can extract a \textit{DNS query event} from the same data. Filters are discussed in detail in subsection \ref{ss:flowanalytics}.

\subsubsection{Enrich sub-module}

A particular enrich sub-module can enrich an attribute with related data. For example, we can enrich an URL with the host address. As before, multiple enrichment can be done on the same piece of data.

\subsubsection{Malicious Scoring sub-module}

This is a specialized sub-module that assigns a malicious score to each piece of data and periodically updates the scores. The novel scoring scheme is discussed in section \ref{sec:pagerank}.

\subsubsection{Automated Defensive Rule Generation sub-module}\label{sss:rule}

This sub-module automatically generates defensive rules (e.g. firewall or intrusion detection system rules) based on the malicious score of the attributes. The rules are published as a feed for users to subscribe.


\subsubsection{Phishing URL Detection sub-module}

This is another specialized sub-module that automatically detects phishing URLs. We have trained a machine learning classifier with features of many labeled URLs. This sub-module is further described in section \ref{sec:usecase}.

\subsection{Report Module}\label{subsec:report}

CYBEX-P is unique in storing cyberthreat data as graphs where the vertices are attributes (e.g. an IP) or events (e.g. an email). This allows CYBEX-P to correlate the data and generate insightful reports. Here, we briefly introduce the report module (\inlinefig{4}, \inlinefig{5}, \inlinefig{6.2}, \inlinefig{7}, \inlinefig{8}, \inlinefig{9}, \inlinefig{10} in Fig. \ref{fig:sysarch}).

Users request reports via the frontend client (\inlinefig{10},\inlinefig{9}). The API (\inlinefig{4}) stores the requests in the cache data lake (\inlinefig{5}). The report server (\inlinefig{6.2}) handles those requests by getting relevant data from the archive DB (\inlinefig{7}) and aggregating them into reports. It then stores the reports in the report DB (\inlinefig{8}). Users can access the reports on demand.

The incident investigation tool, described in subsubsection \ref{sss:neo4j}, is also part of the report module. It provides an interactive graph to explore relationships between different attributes and events.

\section{Data Flow through Entire Lifecycle}\label{sec:dataflow}

Fig. \ref{fig:overview} shows the four major functions of CYBEX-P --- (1) Data collection, (2) Data analysis, (3) Privacy Preservation, and (4) Reporting. This section demonstrates how CYBEX-P achieves these functionalities by following the entire flow of cyberthreat data through it. 

\subsection{Data Input}

Users can manually upload threat data, like a spam email, through the frontend webapp (\inlinefig{10},\inlinefig{9}). Users can also configure the collector (\inlinefig{3}) to automatically collect data from machines like firewalls. Manually uploaded data are directly posted to the API (\inlinefig{4}) whereas automatically collected data are handled by the collector.

\subsection{Privacy Configuration of Data}

The frontend further allows the user to attach an ACL to each piece of data. The ACL dictates which attributes are encrypted. The encryption is done at the connector (\scalerel*{\includegraphics{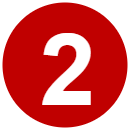}}{|}) or the webapp client (\inlinefig{10}) both of which are at the client premises. 

Although, CYBEX-P cannot access the encrypted attributes, it can still correlate them to generate reports. Users can also share the encryption keys with trusted people. \ref{ss:fcorr} and \ref{sec:datagov} explain the novel correlation mechanism  and the encryption scheme respectively.

\subsection{Data Collection}\label{ss:datacoll}

\begin{figure}[ht]
	\includegraphics[height=.81in]{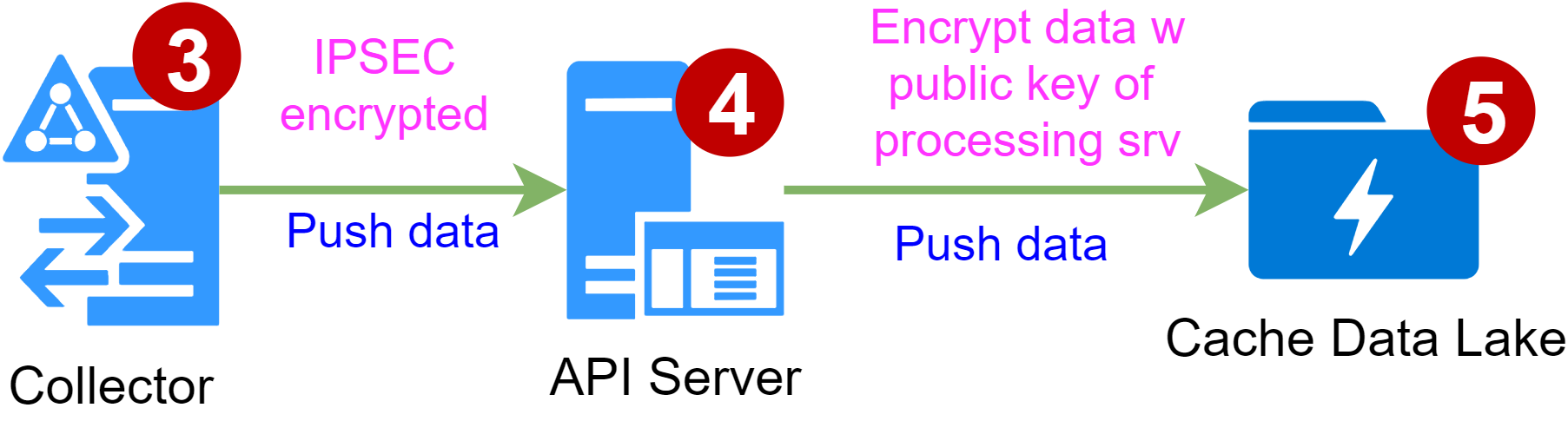} 
	\centering
	\caption{Data Collection in CYBEX-P.}
	\label{fig:datacollection}
\end{figure}

Automatically collected data is forwarded to the collector (\scalerel*{\includegraphics{3}}{|}) over an encrypted channel (TLS). Afterwards, it posts the data to the API (\scalerel*{\includegraphics{4}}{|}). The API encrypts the data with the public key of the archive cluster (\scalerel*{\includegraphics{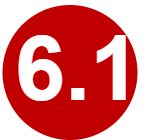}}{|}) and stores the encrypted data in the cache data lake (\scalerel*{\includegraphics{5}}{|}).

The cache data lake acts like a queue or buffer for all incoming data. It also increases security by removing the need for API and archive cluster to communicate directly.

We encrypt the cache data lake with the public key of the archive cluster as the cache data lake is in the demilitarized zone (DMZ). So, the data remains secured even if all the servers in DMZ get compromised.

\subsection{Data Archiving}

\begin{figure}[ht]
	\includegraphics[height=.99in]{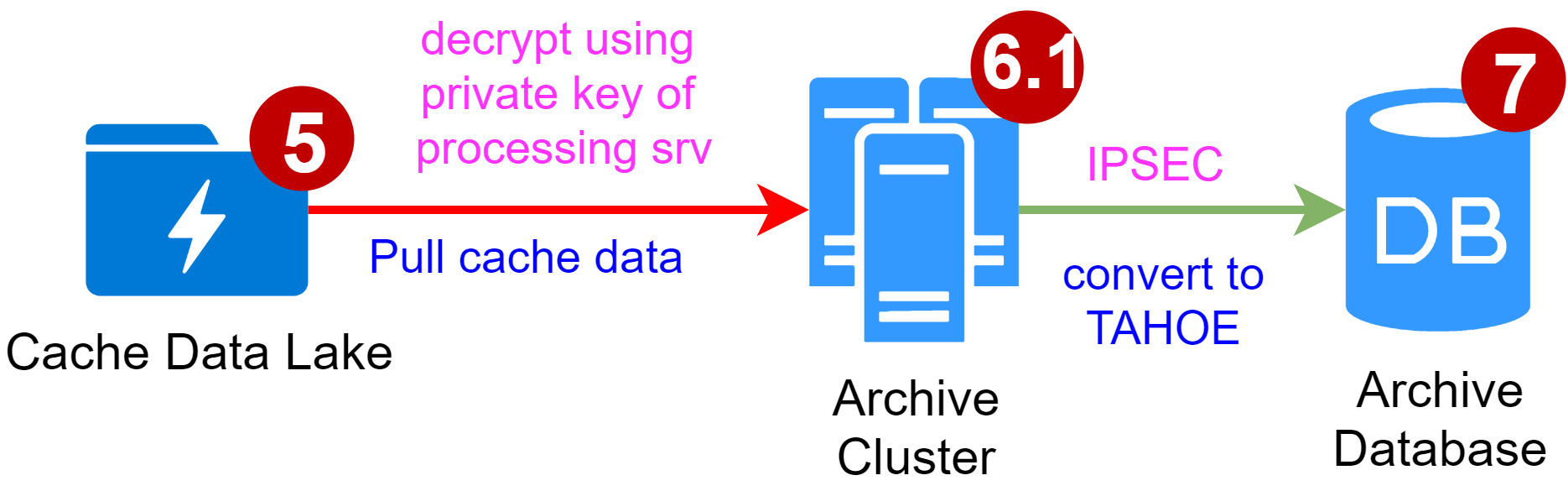} 
	\centering
	\caption{Data Archiving in CYBEX-P.}
	\label{fig:dataarchiving}
\end{figure}

The archive cluster (\scalerel*{\includegraphics{6.1}}{|}), then pulls the data from the cache data lake (\scalerel*{\includegraphics{5}}{|}), decrypts the data using its private key, converts them to TAHOE format and stores them in the archive database (\scalerel*{\includegraphics{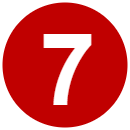}}{|}). All further anlaysis are performed on TAHOE data. TAHOE is discussed in detail in \ref{sec:tahoe}.

\subsection{Data Analytics}\label{ss:flowanalytics}

The analytics cluster (\scalerel*{\includegraphics{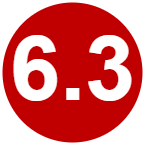}}{|}) transforms, analyzes and correlates data. It achieves that by reading data from the archive database (\scalerel*{\includegraphics{7}}{|}), processing the data in the analytics cluster (\scalerel*{\includegraphics{6.3}}{|}) and writing the processed the data back in the archive database.

\begin{figure}[ht]
	\includegraphics[height=.99in]{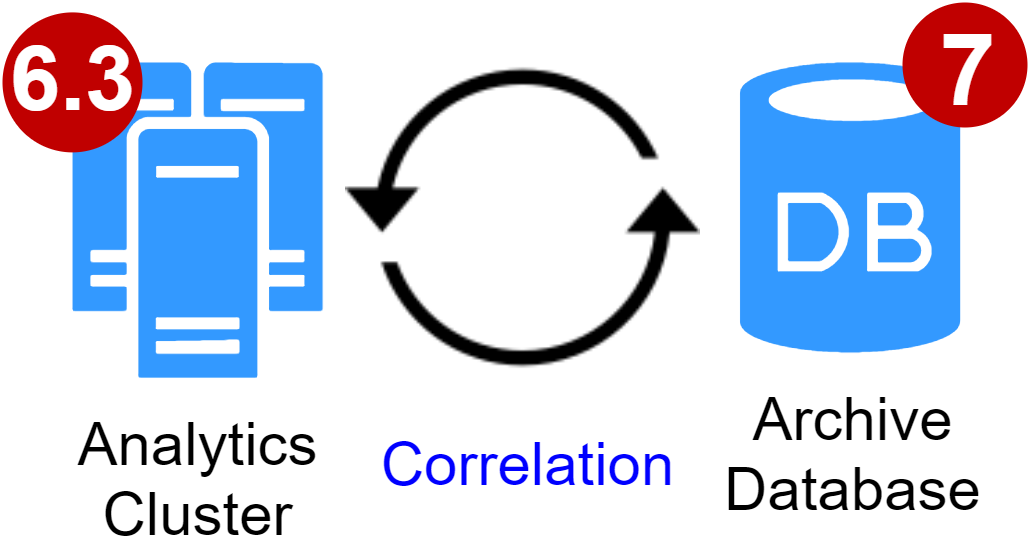} 
	\centering
	\caption{Data Analytics in CYBEX-P.}
	\label{fig:datanalytics}
\end{figure}

This is a continuous process as highlighted by a pair of circular arrows between \scalerel*{\includegraphics{6.3}}{|} and \scalerel*{\includegraphics{7}}{|} in Fig. \ref{fig:datanalytics}. It also is the basis for data correlation in CYBEX-P. 

\begin{figure}[!ht]
	\includegraphics[height=1.5in]{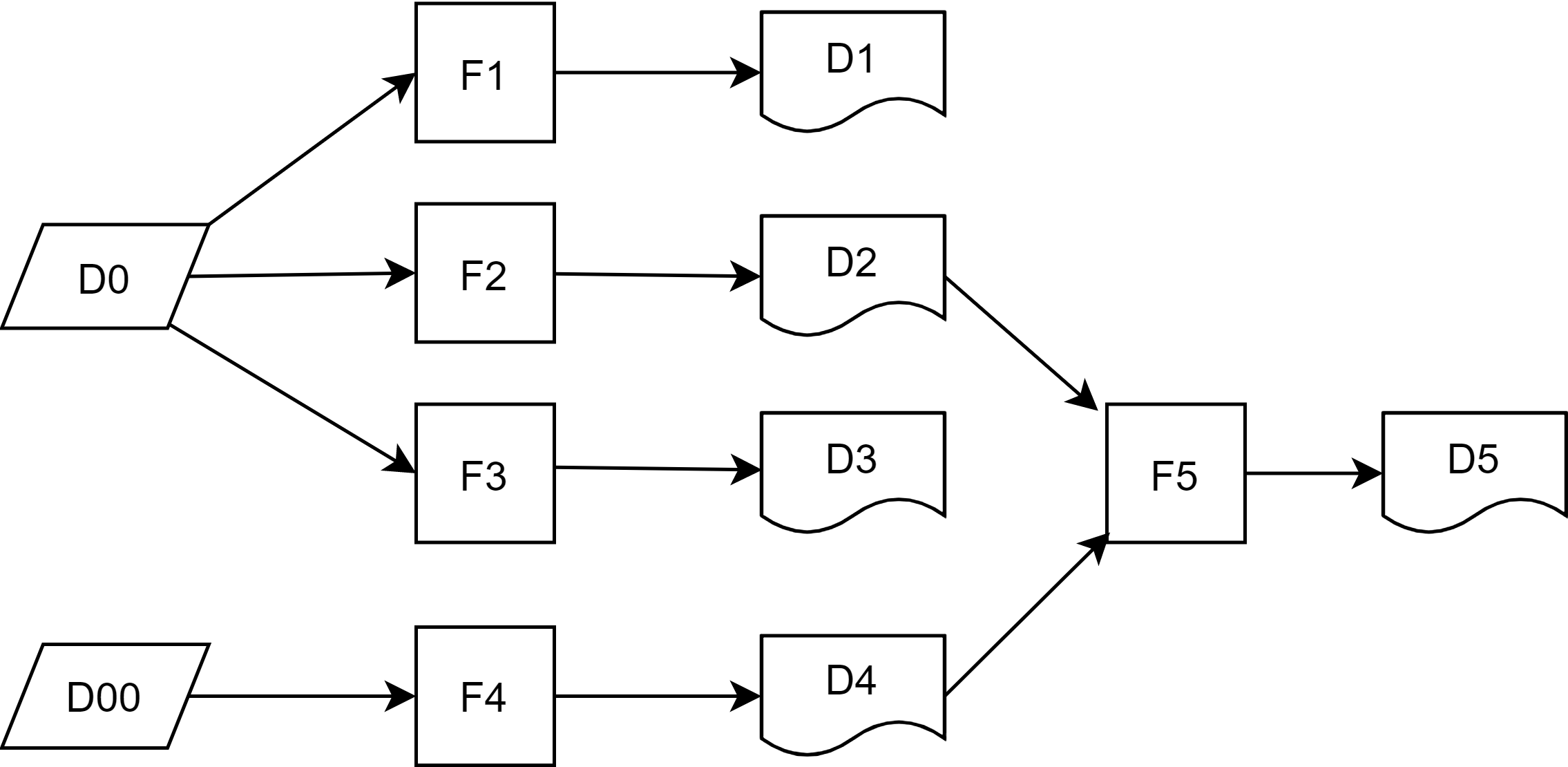}
	\centering
	\caption{Data Filtering as a continuous process.}
	\label{fig:datafiltering}
\end{figure}

For example, consider Fig. \ref{fig:datafiltering} where $3$ filters $F1, F2, F3$ act on a data $D0$ to produce $D1, D2, D3$. An example of such filtering is extracting the source IP, destination IP and destination port from a firewall log. 

Now, $D2$ is further filtered by $F5$ to create $D5$. On the other hand, $D00$ passes through $F4$ and $F5$ to produce the same attribute $D5$. As a direct consequence of how TAHOE works, $D0$ and $D00$ are now connected to each other in a graph via $D5$. This is a very powerful notion in TAHOE, and we use this to assign malicious scores to new events in \ref{sec:pagerank}.


\subsection{Data Reporting}

\begin{figure}[!ht]
	\includegraphics[height=2.2in]{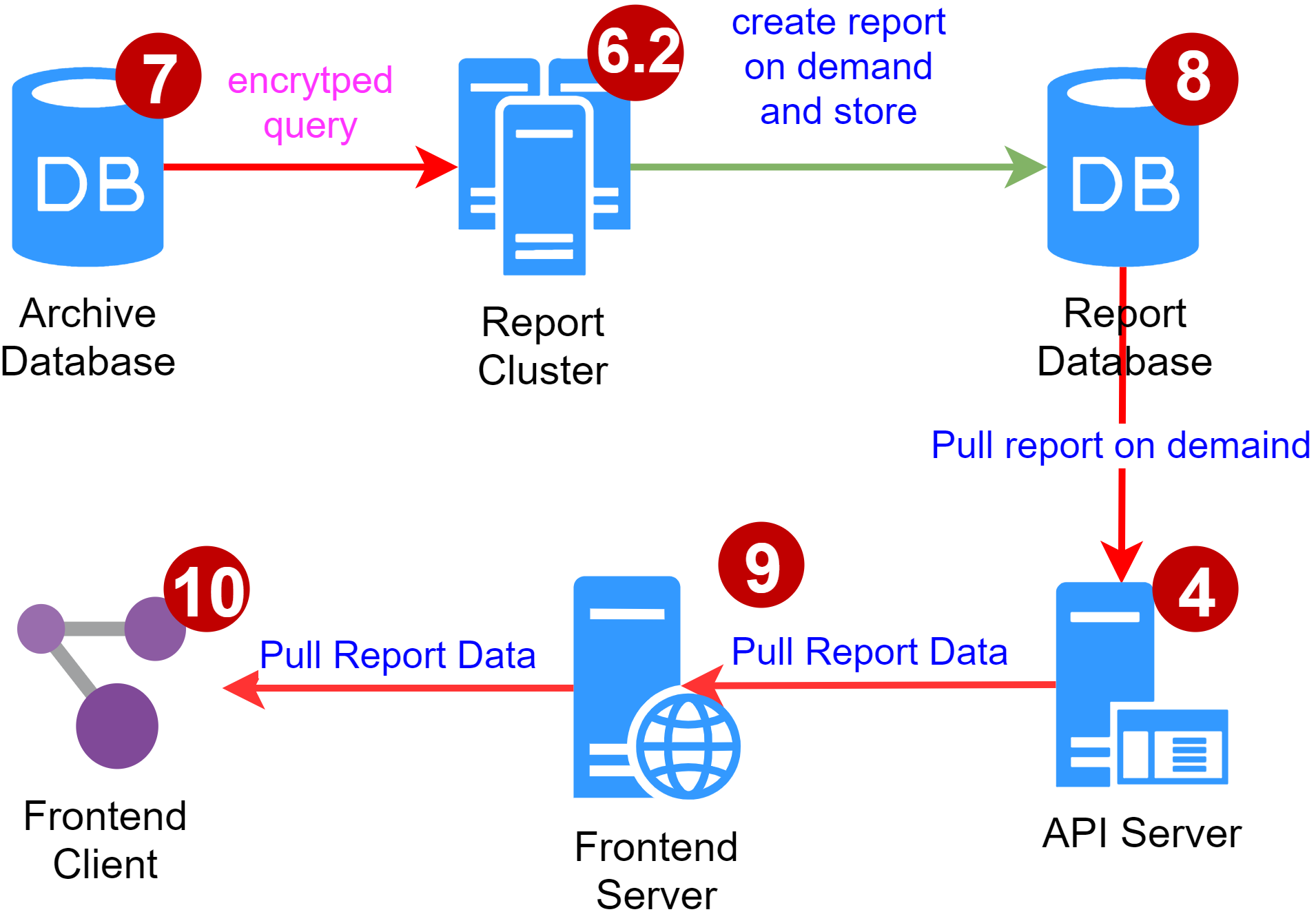} 
	\centering
	\caption{Data Reporting in CYBEX-P.}
	\label{fig:datareporting}
\end{figure}

The data pipeline for requesting a report to CYBEX-P is: User $\Rightarrow$ Frontend client (\scalerel*{\includegraphics{10}}{|}) $\Rightarrow$ frontend server (\scalerel*{\includegraphics{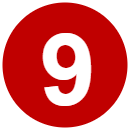}}{|}) $\Rightarrow$ API (\scalerel*{\includegraphics{4}}{|}) $\Rightarrow$ Cache data lake (\inlinefig{5}) $\Rightarrow$ Report cluster (\scalerel*{\includegraphics{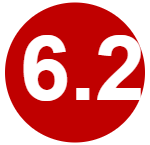}}{|}).

A report can be as simple as counting the occurrence of a particular IP address within a specific time range; on the other hand a report can be as complex as analyzing the attributes of an URL to determine if it is malicious or benign. Nevertheless, the report cluster stores all the reports in the report database. The user can access the reports as follows: User $\Rightarrow$ frontend (\inlinefig{9}, \inlinefig{10}) $\Rightarrow$ API (\scalerel*{\includegraphics{4}}{|}) $\Rightarrow$ Report database (\scalerel*{\includegraphics{8}}{|}).

Note that, the overall process of requesting and getting a report is asynchronous.

\section{TAHOE --- A Cyberthreat Language}\label{sec:tahoe}

TAHOE is a cyberthreat language (CTL). It structures threat data as JavaScript Object Notation (JSON) \cite{bray2014javascript} documents.

Earlier version of CYBEX-P used STIX \cite{barnum2012standardizing} and MISP \cite{wagner2016misp}, presently two of the most popular CTLs. In this section, we introduce TAHOE as a better alternative to traditional CTLs like STIX and MISP.

\subsection{TAHOE Data Instance}\label{ss:instance}

A piece of TAHOE data is called an \texttt{instance} and there are $5$ types of TAHOE \texttt{instances} ---

\subsubsection{\textbf{Raw}} A \texttt{raw} data instance stores unprocessed user data.
\subsubsection{\textbf{Attribute}} The most basic datatype that holds a single piece of information, like an IP address.
\subsubsection{\textbf{Object}} Groups several \texttt{attributes} together, e.g., a file \texttt{object} may have a filename and a size \texttt{attribute}.
\subsubsection{\textbf{Event}} An \texttt{event} consists of one or more \texttt{attributes} or \texttt{objects} along with a \texttt{timestamp}. \texttt{Events} structure \texttt{attributes} or \texttt{objects} into complete threat data (e.g. an email) .
\subsubsection{\textbf{Session}} A \texttt{session} keeps track of arbitrarily related \texttt{events} (e.g. events recorded when a user visits a website).

\subsection{Data Structured as Graphs}\label{ss:graph}
TAHOE structures data as graphs, where the nodes are \texttt{instances}. This structure is explained below.

\begin{figure}[!ht]
	\includegraphics[height=.9in]{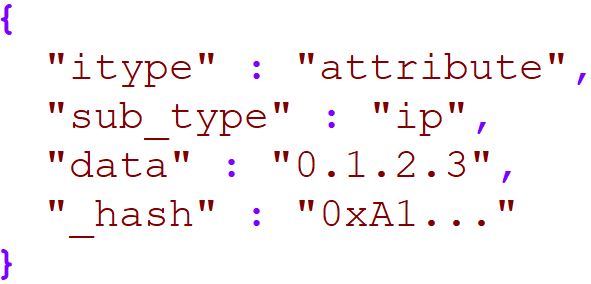} 
	\centering
	\caption{A TAHOE \texttt{ip Attribute}; \texttt{\_hash} also serves as id.}
	\label{fig:0xa1}
\end{figure}

Fig. \ref{fig:0xa1} shows a TAHOE \texttt{ip attribute}. Each TAHOE \texttt{instance} (\texttt{attribute}, \texttt{object}, \texttt{event} etc.) has a \textbf{\texttt{\_hash}} property calculated as the the SHA256 \cite{dworkin2015sha} digest of all other fields and serves as the unique \textbf{id} for that instance. Here, for example, it is the SHA256 of the $<$\texttt{itype,sub\_type,data}$>$ tuple (truncated).

\begin{figure}[!ht]
	\includegraphics[height=2.70in]{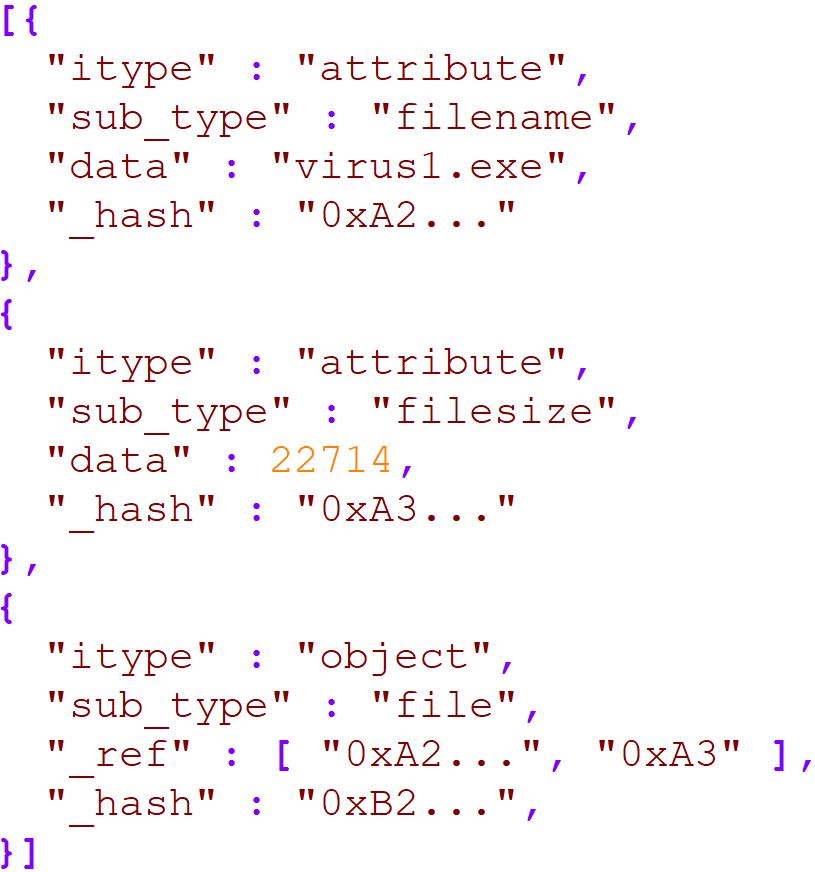} 
	\centering
	\caption{A TAHOE \texttt{file Object} as a graph.}
	\label{fig:0xb1}
\end{figure}

Fig. \ref{fig:0xb1} shows a TAHOE \texttt{file object} with $2$ \texttt{attributes}. As seen, the \texttt{file} object does not include the actual \texttt{filename} and \texttt{filesize} (bytes), rather references them in the \textbf{\texttt{\_ref}} field. So, its complete representation includes these \texttt{attributes} in the outermost array. \texttt{Objects} can also refer other \texttt{objects} if required.

The \texttt{\_ref} field essentially makes the data a graph with $3$ nodes (\texttt{0xA2.., 0xA3.., 0xB2..}) and $2$ edges (\texttt{0xA2.. $\gets$ 0xB2.., 0xA3.. $\gets$ 0xB2..}).

\begin{figure}[!ht]
	\includegraphics[height=2.55in]{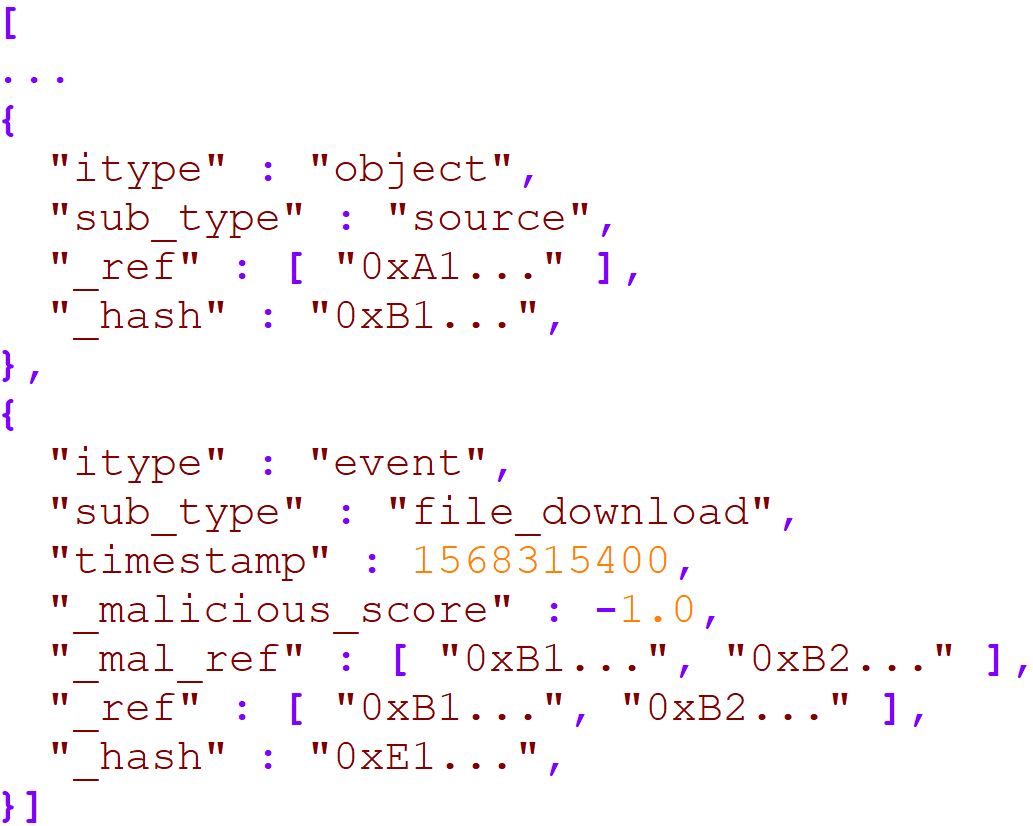} 
	\centering
	\caption{A TAHOE \texttt{file\_download Event} (truncated)\\as a graph of \texttt{objects} and \texttt{attributs}.}
	\label{fig:0xe1}
\end{figure}

Fig. \ref{fig:0xe1} shows a truncated TAHOE \texttt{file\_download event} with $2$ \texttt{objects}. This \texttt{event} refers $2$ \texttt{objects} -- \texttt{0xB1..} (Fig. \ref{fig:0xb1}) and \texttt{0xB2..}. So, both the \texttt{objects} and their \texttt{attributes} must be included in the array (truncated in figure). Events can also refer \texttt{attributes} directly. \texttt{\_malicious\_score} and \texttt{\_mal\_ref} are explained in \ref{sec:pagerank}.

\begin{figure}[!ht]
	\includegraphics[height=1.5in]{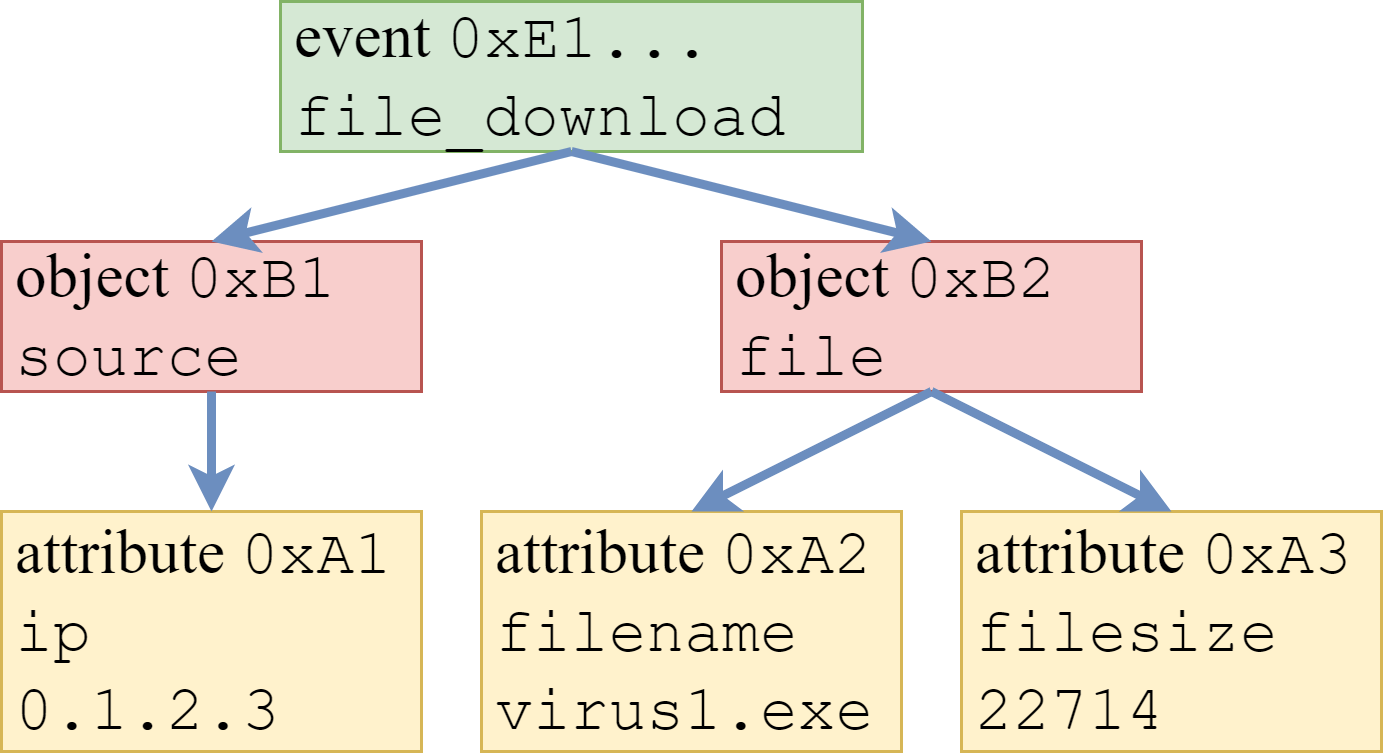} 
	\centering
	\caption{Visual representation of a TAHOE \texttt{event} (green) as a graph of \texttt{objects} (red) and \texttt{attributes} (yellow).}
	\label{fig:graph}
\end{figure}

Figure \ref{fig:graph} shows the visual representation of the event graph with $6$ nodes and $5$ edges. Note, how the \texttt{event} contains complete information on a \texttt{file\_download event} just by referring to a \texttt{source object} and a \texttt{file object}. Benefits of this graphical structure are justified in \ref{ss:fcorr}.

Note that, we draw the edges as arrows because of how edge data is stored in \texttt{\_ref}. However, in TAHOE database, a graph can be traversed from both end, as explained in \ref{sss:biq}.

\begin{figure}[!ht]
	\includegraphics[height=.6in]{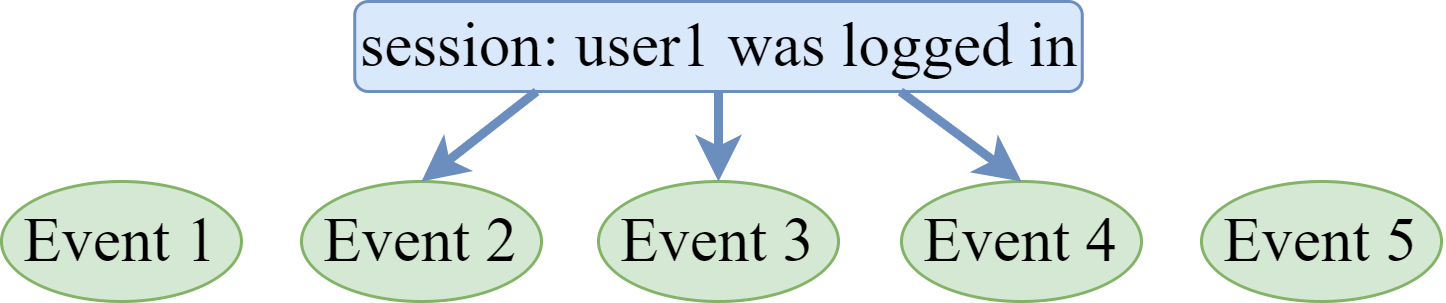} 
	\centering
	\caption{\texttt{Events} grouped by arbitrary session parameter.}
	\label{fig:session}
\end{figure}

Finally, A TAHOE \texttt{session} is an arbitrary grouping of related \texttt{events}. This allows us to group together \texttt{events} based on any condition the user desires. The \texttt{session} in Fig. \ref{fig:session} groups $3$ events, recorded while $user1$ was logged in.

\subsection{Events Viewed as Nested Documents}

\begin{figure}[!h]
    \centering
    \includegraphics[height=1.68in]{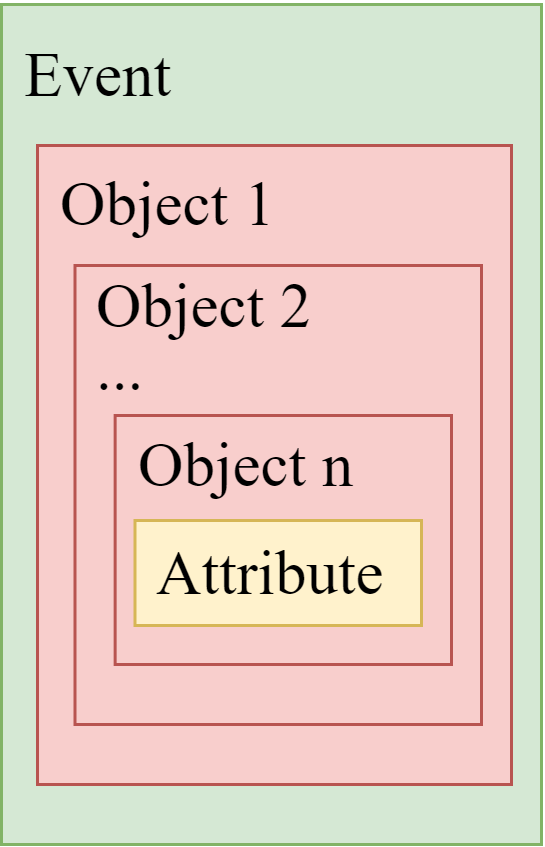}
    \caption{A Generic TAHOE \texttt{event} visualized as a nested document.}
    \label{fig:nestedgeneral}
\end{figure}

\begin{figure}[!h]
    \centering
    \includegraphics[height=1.68in]{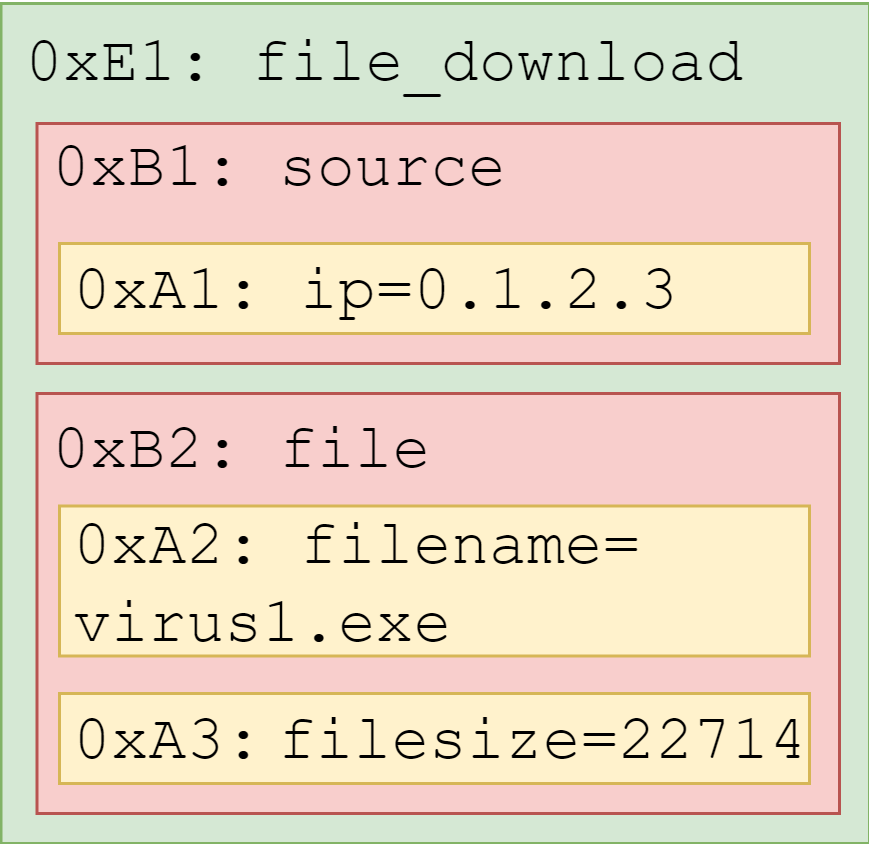}
	\caption{Event \texttt{0xE1..} from Fig.\ref{fig:graph} as a nested document.}
	\label{fig:nestedexample}
\end{figure}

Though, TAHOE structures \texttt{events} as graphs, they can be viewed as nested documents. Fig. \ref{fig:nestedgeneral} shows the general visualization of a TAHOE event as a nested document, while Fig. \ref{fig:nestedexample} shows the event from Fig. \ref{fig:graph} as a nested document.

As described in subsection \ref{ss:graph}, \texttt{objects} can refer other \texttt{objects}. That means \texttt{objects} can be infinitely nested as presented in \ref{fig:nestedgeneral}. This allows TAHOE to store arbitrarily complex data.

An analyst can choose to view an \texttt{event} as a document or as a graph depending on her need. For all kinds of machine analysis (e.g query), however, the graphical structure of Fig. \ref{fig:graph} is more suitable.

\subsection{Intrinsic Correlation of Graphical Data}\label{ss:fcorr} 

\begin{figure}[!ht]
	\includegraphics[height=.75in]{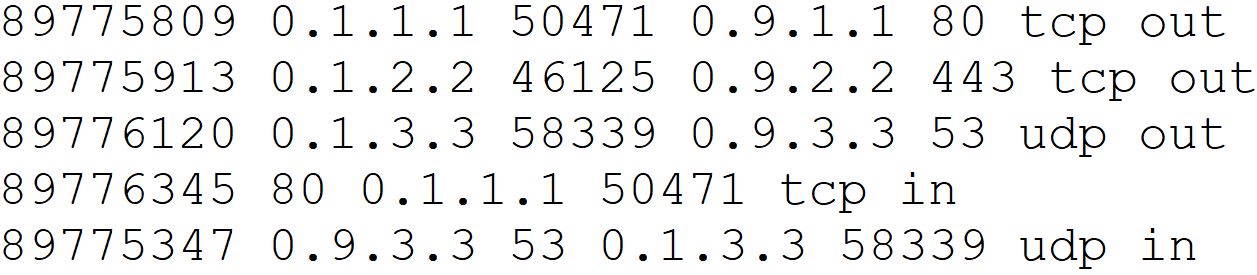} 
	\centering
	\caption{Firewall log (as documents) is a clutter.}
	\label{fig:samplelog}
\end{figure}

Traditional CTLs, like STIX and MISP, store threat data as documents. Fig. \ref{fig:samplelog} shows part of a typical firewall log. Each row is an \texttt{event} with $7$ \texttt{attributes}; traditionally stored as a document. It is cumbersome to analyze because the \texttt{events} lack any direct correlation with own \texttt{attributes}.

\begin{figure}[!ht]
	\includegraphics[height=1.86in]{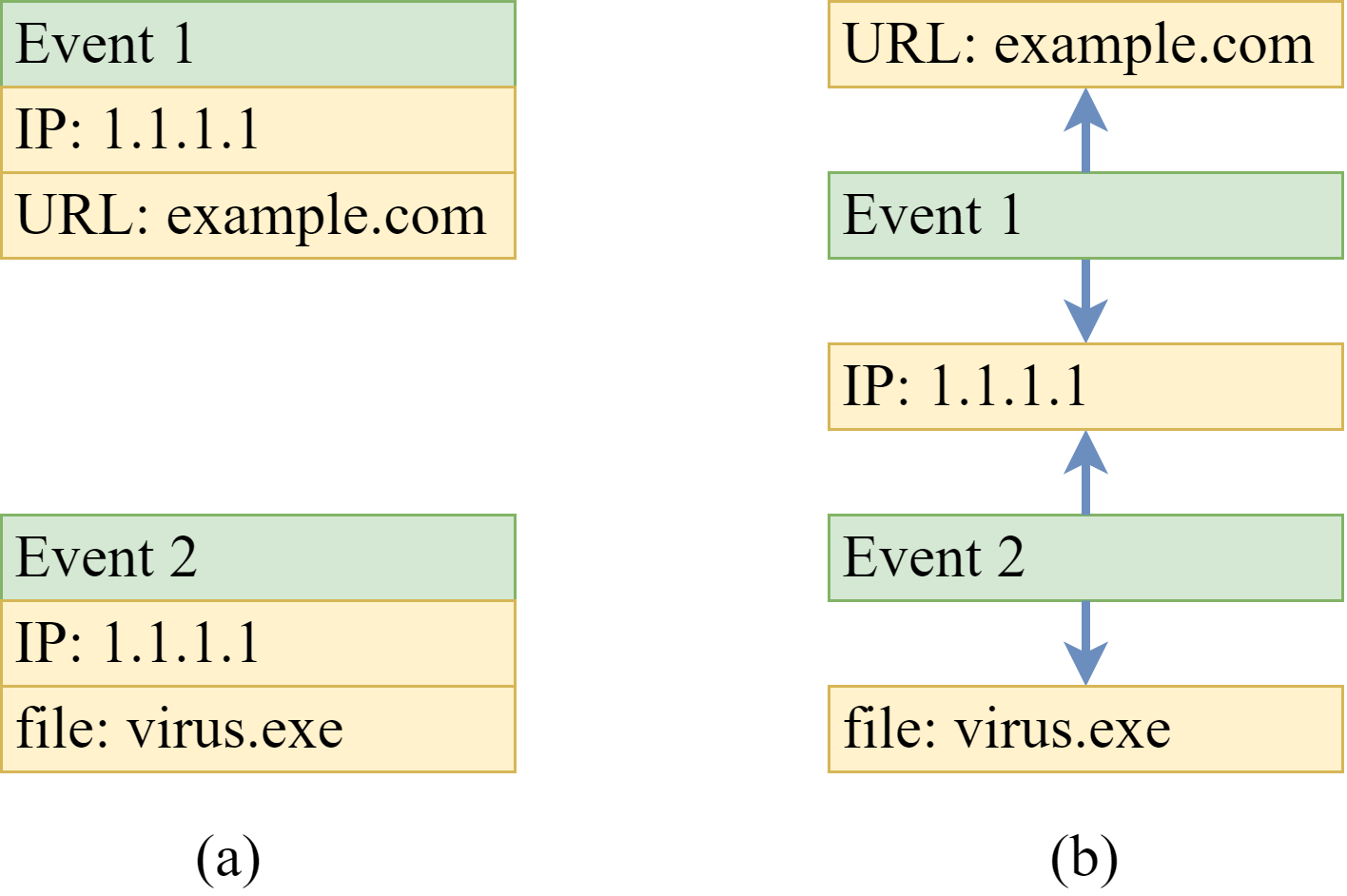} 
	\centering
	\caption{Transforming two \texttt{events} from (a) separate documents to (b) TAHOE graph reveals apparent correlation.}
	\label{fig:tahoe}
\end{figure}

TAHOE, on the other hand, represents data as graphs like in Fig. \ref{fig:tahoe}. Here, two separate \texttt{events} are automatically connected by their common \texttt{attribute} (\texttt{1.1.1.1}) in TAHOE.

Such `intrinsic correlation' is a powerful feature of TAHOE, because if someone looks up \texttt{example.com} she will immediately see that \texttt{virus.exe} is related to it. This is a major strength of our investigation tool (subsubsection \ref{sss:neo4j}). Moreover, we leverage this feature to formulate a novel malicious event detection mechanism in section \ref{sec:pagerank}.

\subsection{Representing Complex Data \& Supporting Fast Query}\label{ss:complex}

\begin{figure}[!ht]
	\includegraphics[height=.72in]{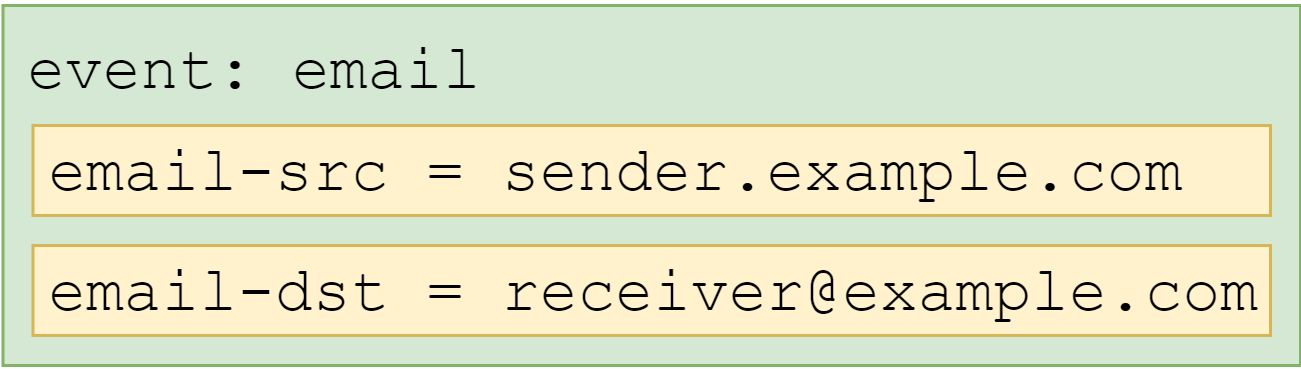} 
	\centering
	\caption{MISP stores data in $2$ layers for fast queries but has to create non-intuitive \texttt{attribute} types as a result.}
	\label{fig:misp}
\end{figure}

Traditional CTLs make a trade-off between representing complex data and fast query. For example, MISP stores data in a relational database (RDBMS) for fast queries. However, RDBMSs do not support arbitrary data structures. So, MISP structures data in only $2$ layers -- \texttt{attribute} \& \texttt{event}, as shown in Fig. \ref{fig:misp}.

The problem is, to fetch all emails to and from \texttt{doe@example.com}, one has to perform $2$ queries -- \texttt{email-src = doe@example.com} and \texttt{email-dst = doe@example.com}. For the same reason, MISP has to specify cumbersome \texttt{attribute} types like \texttt{passenger-name-record-locator-number} or non-intuitive \texttt{attribute} types like \texttt{filename|md5, filename|sha224, filename|sha256} etc.



On the other end of the spectrum, we have TAXII \cite{connolly2014trusted}, which structures data as STIX. TAXII by default stores data in a NoSQL database (MongoDB), to support arbitrary data structures. However, there are hundreds of threat data types (IP, email address, URL etc.), while only $64$ types can be indexed in MongoDB. Querying un-indexed data takes more than hours in a reasonably sized TAXII server, making it practically unfeasible.

\begin{figure}[!ht]
	\includegraphics[height=.52in]{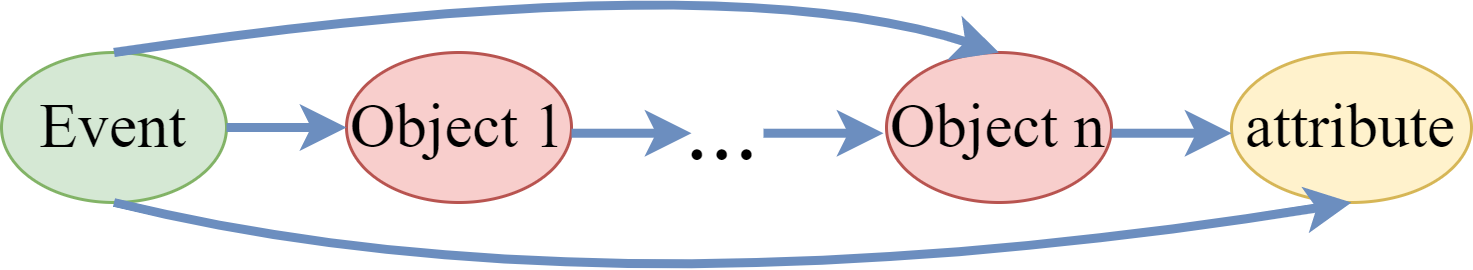} 
	\centering
	\caption{A TAHOE \texttt{event} refers all subsequent nodes.}
	\label{fig:tahoeref1}
\end{figure}

TAHOE tackles these challenges by employing a set of novel techniques. Firstly, TAHOE can store arbitrarily complex data because TAHOE \texttt{objects} can refer other \texttt{objects}, as depicted by $n$ \texttt{objects} in Fig. \ref{fig:tahoeref1}. In other words, TAHOE \texttt{objects} are infinitely nested, as shown in Fig. \ref{fig:nestedgeneral}.

Secondly, a TAHOE \texttt{event} refers all subsequent \texttt{nodes} up to the leaf, not only the next node. This is depicted in Fig. \ref{fig:tahoeref1}, by the edges -- \texttt{event} $\rightarrow$ \texttt{object n} and \texttt{event} $\rightarrow$ \texttt{attribute}. As a result, any TAHOE \texttt{event} is always $1$ database query away from it's \texttt{attributes} no matter the level of nesting.

Finally, TAHOE databases do not have to index all \texttt{attribute} types, rather just $2$ fields -- (1) the \texttt{\_hash} field to lookup an \texttt{attribute} and the \texttt{\_ref} to lookup related \texttt{events}. This is explained in detail in \ref{sss:idx}.

\begin{figure}[!ht]
	\includegraphics[height=1.8in]{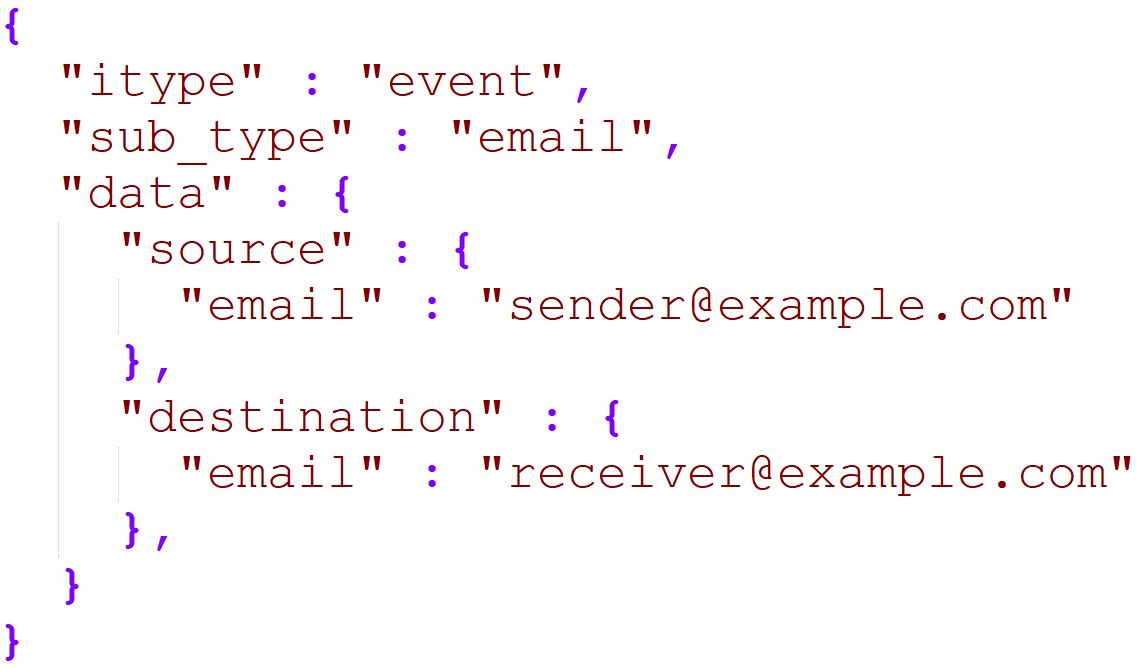} 
	\centering
	\caption{A TAHOE \texttt{email event} as a nested document.}
	\label{fig:emaildoc}
\end{figure}

\begin{figure}[!ht]
	\includegraphics[height=1.48in]{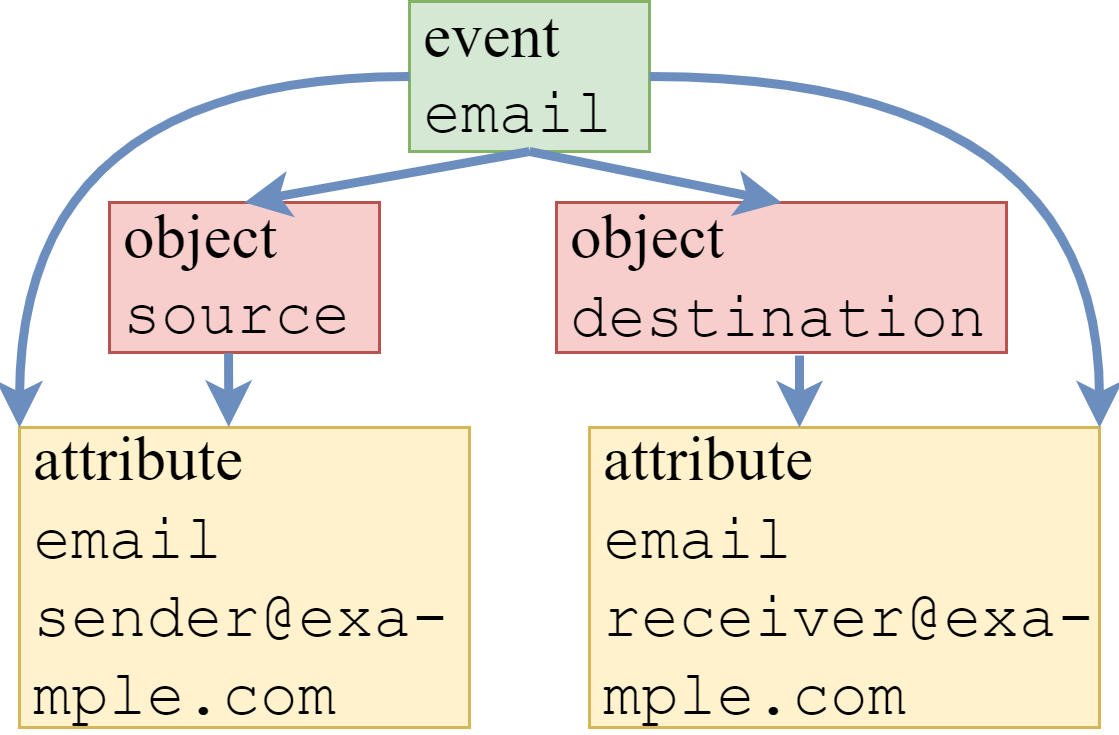} 
	\centering
	\caption{A TAHOE \texttt{email event} as a graph.}
	\label{fig:emailgraph}
\end{figure}

Now, going back to the example of Fig. \ref{fig:misp}, we can structure an \texttt{email} in TAHOE as Figures \ref{fig:emaildoc} and \ref{fig:emailgraph}. Note that both the source and destination email addresses are stored in TAHOE \texttt{email attributes}, not two different \texttt{attributes} like MISP. So, fetching all emails to and from \texttt{doe@example.com} will take only one query. Moreover, as all the \texttt{events} connected to \texttt{doe@example.com} are zero hop away, the query is fast.

\subsection{Threat Data Query Language (TDQL)}\label{ss:tdql}

TAHOE aims to standardize the structuring of threat data in terms of \texttt{attributes, objects, events} and \texttt{sessions}. This would allow users to query threat data using those terms. An example query could be \texttt{fetch all events which include the attribute 1.1.1.1}. At present this is not possible because \texttt{event} or \texttt{attribute} are not standardized terms for any existing database. For example, if a person queries an SQL database for \texttt{events} it would not know what to return, because \texttt{event} is not a standard term for SQL.


To that end, we have developed a universal threat data query language (TDQL) for TAHOE. TDQL acts as a layer between a database and a user.  Additionally, TDQL is tailor made for threat data and addresses their nuances. While SQL depends on the structure of database tables, TDQL speaks in terms of \texttt{attributes, objects, events} etc. So, irrespective of the data storage or delivery protocol, a user can always fetch any threat data from any database.

Additionally, having a dedicated TDQL makes TAHOE, database-independent. However, detailed documentation of TDQL is beyond the scope of this research work.

\subsection{Features of TAHOE}

\subsubsection{\textbf{Data Normalization}} TAHOE normalizes different formats of same type of data. Consider two firewalls from two different vendors. Their log data will be formatted differently despite having same type of data. TAHOE normalizes such differences by converting them into the same structure.

\subsubsection{\textbf{Data De-duplication}}\label{sss:dup}

TAHOE prohibits duplicate data. For example, there can only be one instance of the IP $1.1.1.1$ in a TAHOE database. This saves CYBEX-P a lot of storage by not storing the same IP in different \texttt{events}. TAHOE achieves this de-duplication of data by creating a globally reproducible hash of the data.

\begin{figure}[ht]
	\includegraphics[height=.91in]{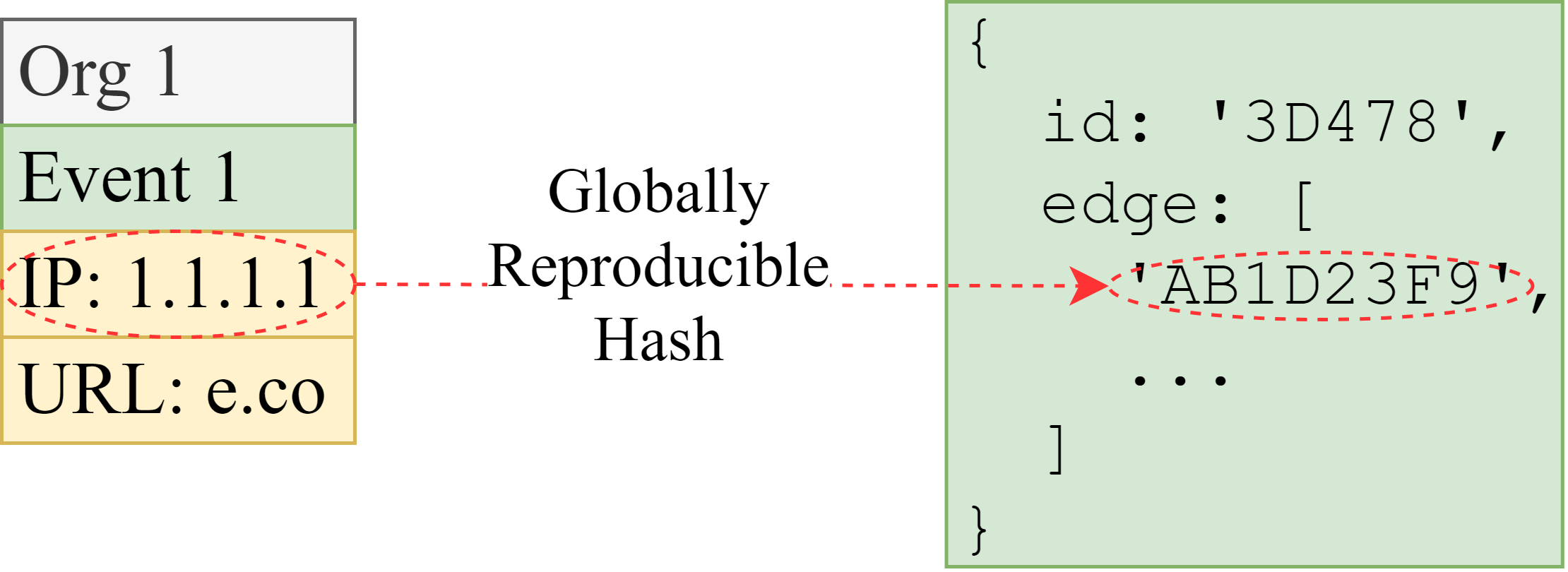}
	\centering
	\caption{Edges are stored as hashes of child document.}
	\label{fig:edge}
\end{figure}

\subsubsection{\textbf{Database Independence}}


Although TAHOE is a graph based CTL we did not use a graph database as a container for TAHOE. In other words, all the information, including the edge data, of a TAHOE graph is stored in the JSON documents of the TAHOE \texttt{instances}, as shown in Fig. \ref{fig:edge}.

Furthermore, as described in \ref{ss:tdql} we have developed a universal threat data query language (TDQL) to communicate with any TAHOE storage. These two contributions make TAHOE a database-independent CTL.

\subsubsection{\textbf{Optimized for Indexing}}\label{sss:idx}

Consider a query to find all \texttt{events} that include a particular IP. Traditionally, one has to index the IP field of all \texttt{events}. This creates $3$ problems ---

\begin{enumerate}
	\item Not all \texttt{events} has the IP field, so the indexing will be inefficient.
	\item There are hundreds of \texttt{attribute} types (e.g. URL, email, IP etc.), all of which cannot be indexed. For example, only $64$ fields can be indexed in MongoDB.
	\item Some \texttt{attributes} have very large values. For example, a `comment' \texttt{attribute} can be larger than $1024$ bytes which is the limit placed by MongoDB.
\end{enumerate}

In TAHOE, however, we would first query the IP node using its \texttt{\_hash}. Next, we would get the related \texttt{events} by querying the edge array (\texttt{\_ref}) of all \texttt{events}. Consequently, we only need to index the edges and \texttt{\_hash} fields, not the actual \texttt{attribute} values. Since both the \texttt{\_hash} and \texttt{\_ref} fields are $256$ bits long, indexing them does not violate any database limits.

\begin{figure}[ht]
	\includegraphics[height=1.96in]{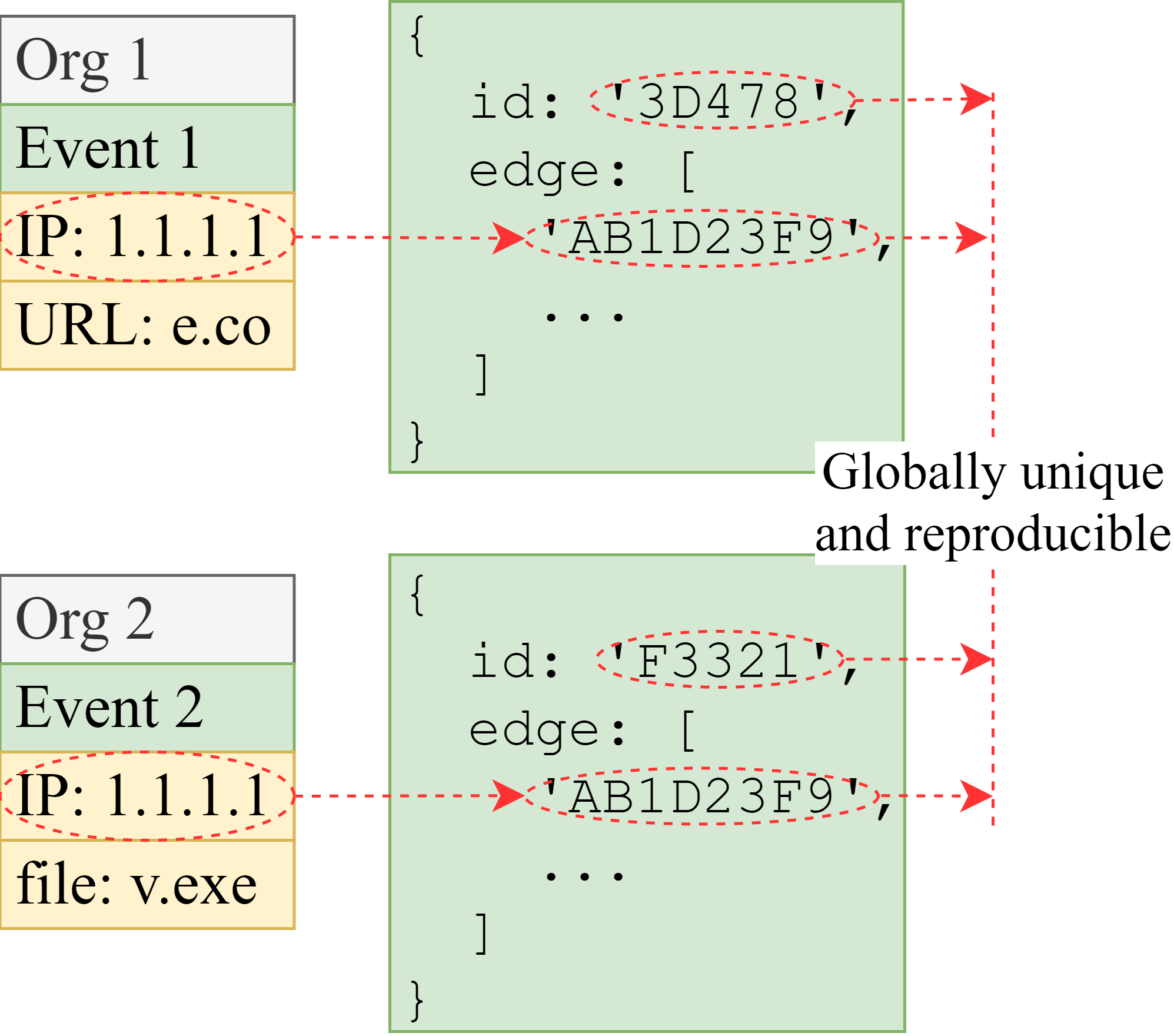}
	\centering
	\caption{TAHOE id and edges are globally unique and reproducible, making them collision free.}
	\label{fig:repedge}
\end{figure}

\subsubsection{\textbf{Globally Unique \& Reproducible Data for Conflict-free Sharing}}

TAHOE data are globally unique and reproducible. As shown, in Fig. \ref{fig:repedge}, the IP \texttt{1.1.1.1} has the same unique id (its hash) in two different organizations. Consider, \texttt{Org 1} shares \texttt{Event 1} with \texttt{Org 2}. If \texttt{Org 2} had a different id for \texttt{1.1.1.1} it would have to update the edge array of \texttt{Event 1}. But, as hashes are reproducible yet unique, this is not required.

Note that, \texttt{event} hashes include a timestamp. As a result, two separate \texttt{events} will have different hashes even if they have the same attributes.

\subsubsection{\textbf{Bidirectional Edges for Versatile Queries}}\label{sss:biq}

TAHOE edges are bidirectional. As seen in Fig. \ref{fig:repedge}, edge data is stored in the \texttt{event} only. This is because, an IP like $8.8.8.8$ (public DNS) can potentially get connected to millions of \texttt{events}. If we store the hash of all these \texttt{events} in the IP \texttt{attribute}, it would result in an unbounded growth of its edge array. So, we store the edge info in the \texttt{events}. However, it takes only one pass over the database, to get all \texttt{events} that have a particular hash in their edge array. So, the edges are bidirectional for all intents and purposes.

\section{ThreatRank to Detect Malicious Events}\label{sec:pagerank}


Earlier in subsection \ref{ss:fcorr} we introduced how TAHOE intrinsically correlates data. Here, we extend upon it by formulating  an algorithm, called ThreatRank, to assign a malicious score to each \texttt{event} in a TAHOE database. The score essentially sorts the \texttt{events} from most malicious to least malicious. In \ref{ss:respage} we justify this algorithm with real data.

Consider, $\mathbb{A} = \{a_1$, $a_2$, ..., $a_m\}$ is the set of all \texttt{attributes} and $\mathbb E = {e_1, e_2, ..., e_n}$ is the set of all \texttt{events}. $\mathbb E_{mal} \subseteq \mathbb E$ is the set of known malicious \texttt{events}. We define $\mathbb I_{mal} = \{k ~ | ~ e_k \in \mathbb E_{mal}\}$. We want to determine the ThreatRank (TR) of a new \texttt{event} $e_p$.

We define $w_{i,j} = \{e_i, ..., a_x, e_y, a_z ..., e_p\}$ as the $j^{th}$ path from $e_i$ to $e_p$. Note that, the path encounters \texttt{attributes} and \texttt{events} in an alternating fashion and has distinct nodes.

Then the contribution of $w_{i,j}$ to the ThreatRank of $e_p$ is calculated using the recurrence equation---

\begin{equation}\label{eq:trwij}
	TR_{w_{i,j}}[k] = 0.998^{d_{k-1}}  \times \frac{ TR_{w_{i,j}}[k-1] }{L( w_{i,j}[k-1] )}
\end{equation}

where, $TR_{w_{i,j}}[1] = -1$; $d_k = 0$ for an \texttt{attribute} and for an \texttt{event}, $d_k$ is the number of days passed since the \texttt{event} $e_k$ was recorded; $L(x)$ is the degree of node $x$.

Assume, there are $t_i$ paths from $e_i$ to $e_p$. We define the set $\mathbb W = \{w_{i,j} ~ | ~ i \in \mathbb I_{mal}; j \in [1,t_i]\}$. $\mathbb W$ basically includes all the paths from all known malicious \texttt{events} to the new \texttt{event}. The total ThreatRank of $e_p$ is then calculated as ---

\begin{equation}\label{eq:trep}
	TR(e_p) = \sum_{w \in \mathbb W} TR_{w}[t_i]
\end{equation}

Algorithm \ref{algo:threatrank} lists the pseudocode for ThreatRank. The code is written using TAHOE terminology.

\begin{algorithm}
    \caption{ThreatRank}
    \label{algo:threatrank}
    \DontPrintSemicolon
    \SetAlgoLined

    \SetKwProg{Fn}{Function}{}{end}
    \SetKwInOut{Input}{Input}

    \Input{$\mathbb E, ~ \mathbb E_{mal}$}

    \Fn{getRelated(node)} {
        \If{node.type = ``event"} {return node.\_ref}
        related $\gets$ []    \; 
        \For{event in $\mathbb E$} {
            \If {node in event.\_ref} {related.append(node)}
        }

        return related \;
    }

    \Fn{findPaths(src, dest, currentPath)} {
        \If{src = dest} {return currentPath }

        related $\gets$ getRelated(src) \;
        paths $\gets$ [] \;

        \For{r in related}{
            \If  {r in currentPath} {continue}  
            paths.append(findPaths(r, dest, currentPath+[r])) \;
        }
        return paths \;
    }

    \Fn{threatRankPath(path)} {
        tr $\gets$ $-1$ \;

        \For{node in path} {
            L $\gets$ degree(node) \;
            d $\gets$ 0 \;
            \If{node.type = ``event"} { d $\gets$ node.daysOld }
            tr $\gets$ tr $\times$ 0.998**d $/$ L \;
        }
    return tr \;
    }

    \Fn{threatRank(newEvent)} {
        allPaths $\gets$ [], ~~ TR $\gets$ 0 \;

        \For{event in $\mathbb E_{mal}$} {
            allPaths.append(findPaths(event, newEvent, []))
        }
        \For{path in allPaths} {
            TR $\gets$ TR $+$ threatRankPath(path) \;
        }
        return TR \;
    }

\end{algorithm}

\subsection{Why 0.998?}

We multiply the ThreatRank of each \texttt{event} by $0.998^{d_k} $. Here, $d_k$ is the number of days passed since the \texttt{event} $e_k$ was recorded. The value $0.998$ is chosen such that after $1$ year an \texttt{event} is half as significant ($0.998^{365} = 0.48$) as a recent \texttt{event} ($0.998^0 = 1$). The same \texttt{event} is only one-fourth as significant ($0.998^{730} = 0.23$) after two years.

\subsection{Who Classifies Malicious Events \& Edges?}

Malicious \texttt{events} or edges can be manually classified in three ways --- (1) by CYBEX-P admin after analysis (2) by user voting (3) automatically for some data. For example, an IP that tries to connect to a honeypot, is automatically classified as a malicious IP in this context.



\section{Data Governance \& Privacy Preservation}\label{sec:datagov}

CYBEX-P offers a robust data governance mechanism with granular access control of data. Here, we discuss the `attribute based access control' protocol of CYBEX-P.

\subsection{Data Model and Privacy Parameters}

\begin{figure}[ht]
	\includegraphics[height=.72in]{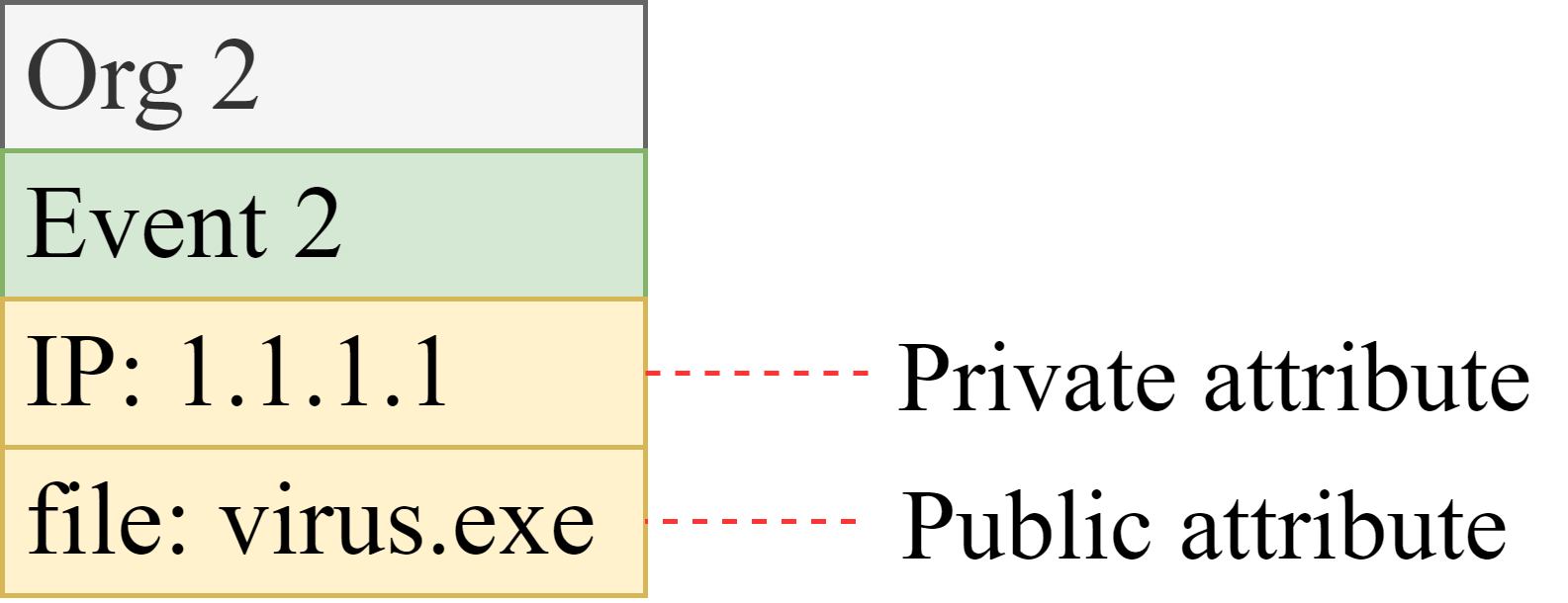} 
	\centering
	\caption{Data owner determines if an attribute is public or private.}
	\label{fig:priv1}
\end{figure}

CYBEX-P converts any incoming data into a TAHOE \texttt{event}.  Fig. \ref{fig:priv1} shows an \texttt{event} with two \texttt{attributes} -- an \texttt{IP} and a \texttt{file} (filename). Assume, the data owner \texttt{Org 2} wants to share the \texttt{file} with everyone but not the \texttt{IP}.

\subsection{Public Attribute (Not Encrypted)}

\begin{figure}[ht]
	\includegraphics[height=.72in]{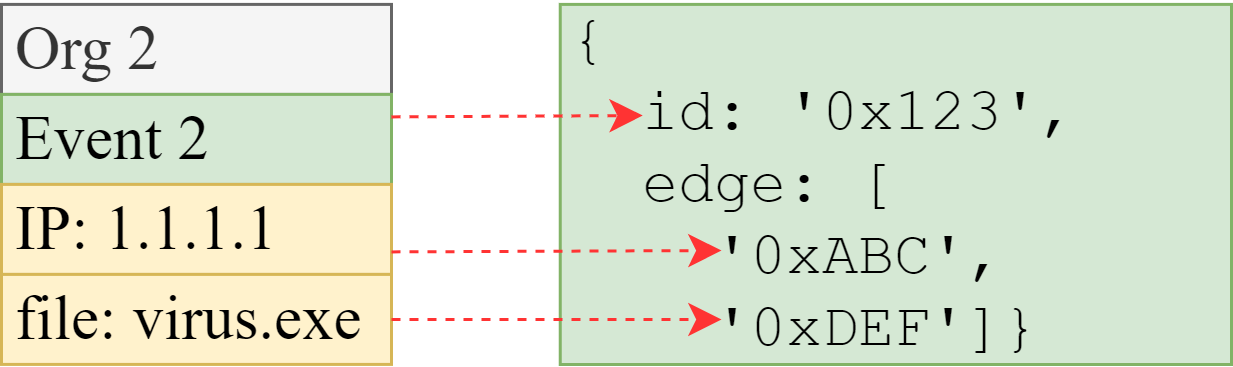} 
	\centering
	\caption{Event data in TAHOE format before encryption.}
	\label{fig:priv2}
\end{figure}

Data owner \texttt{Org 2}, first, converts the document into a TAHOE \texttt{event} as shown in Fig. \ref{fig:priv2}. Here, \texttt{0xABC} is the hash of the IP \texttt{1.1.1.1}, \texttt{0xDEF} is the hash of the file \texttt{virus.exe}, \texttt{0x123} is the hash of the \texttt{event} itself, and acts as the event id. The hashes of the attributes are placed in the \texttt{edge} array creating a graph. Note that, we use the term \texttt{edge} to denote the \texttt{\_ref} array from \ref{ss:graph}.

\subsubsection*{\textbf{Threat Model}}

The data owner \texttt{Org 2} trusts all participants of CYBEX-P with the public attribute \texttt{virus.exe}.

\subsubsection*{\textbf{Public Data Query}}

\begin{figure}[ht]
	\includegraphics[height=.36in]{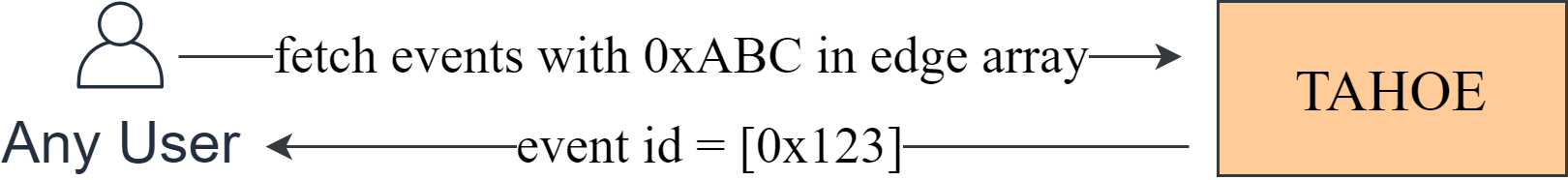} 
	\centering
	\caption{Public Data Query.}
	\label{fig:pubque}
\end{figure}

Now, assume a user wants to get all the events with the file \texttt{virus.exe}. She first generates the hash of \texttt{virus.exe} as \texttt{0xDEF}. Then she looks up the database for events that have \texttt{0xDEF} in the \texttt{edge} array. She will get the \texttt{event 0x123} in return.

\subsection{Private Attribute Encryption}

\begin{figure}[ht]
	\includegraphics[height=1.02in]{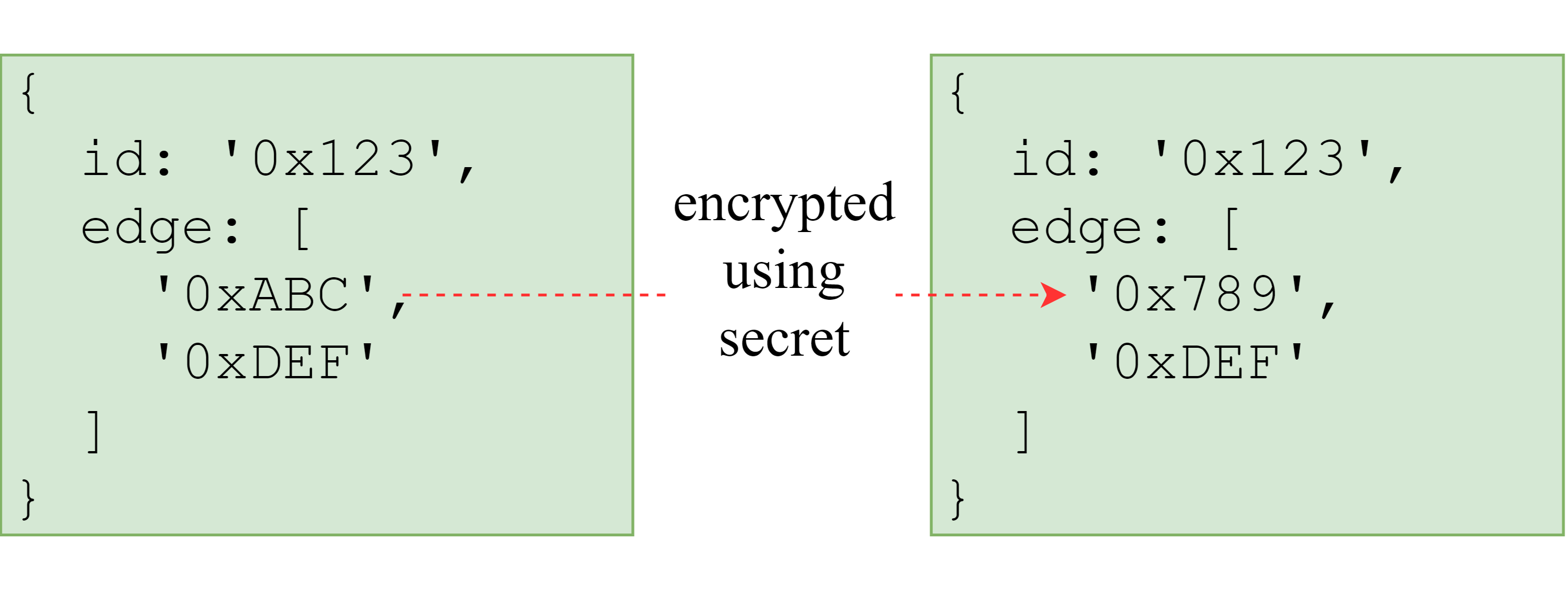} 
	\centering
	\caption{Event data in TAHOE format after encryption.}
	\label{fig:priv3}
\end{figure}

To protect the private attribute, \texttt{Org 2} encrypts its hash \texttt{0xABC} with \texttt{secret} to generate the ciphertext \texttt{0x789}, as shown in Fig. \ref{fig:priv3}. The owner can use any symmetric encryption technique of choice although TAHOE recommends AES256.

\subsubsection*{\textbf{Threat Model}}

The data owner \texttt{Org 2} does not trust anybody including CYBEX-P with the private attribute \texttt{1.1.1.1}.

\subsubsection*{\textbf{Private Data Query}}

\begin{figure}[ht]
	\includegraphics[height=.36in]{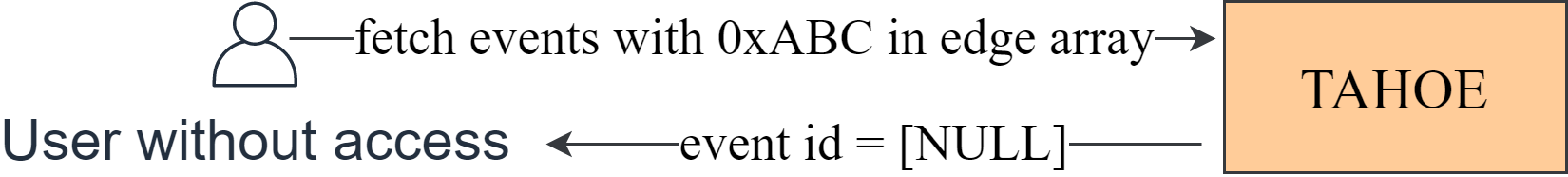} 
	\centering
	\caption{Private Data Query.}
	\label{fig:privque}
\end{figure}

Now, a user wants to fetch all the \texttt{events} with the IP \texttt{1.1.1.1}. She first generates the hash of IP \texttt{1.1.1.1} as \texttt{0xABC}. Then she queries the database for \texttt{events} that have \texttt{0xABC} in their \texttt{edge} array. However, the database will return nothing, because the value \texttt{0xABC} is not present in any event edge. \texttt{Org 2} has essentially encrypted the graph edge.

\subsection{Private Attribute Sharing}

At this point, \texttt{Org 2} wants to share the private attribute \texttt{1.1.1.1} with \texttt{Org 3}. To achieve this, \texttt{Org 2} shares the \texttt{secret} with \texttt{Org 3}. CYBEX-P facilitates this sharing by providing a key management system (KMS) (\inlinefig{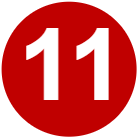} in Fig. \ref{fig:sysarch}).

\subsubsection*{\textbf{Threat Model}}

\texttt{Org 2} trusts \texttt{Org 3} and wants to share \texttt{1.1.1.1} with \texttt{Org 3}. However, \texttt{Org 2} does not trust CYBEX-P or any other user. \texttt{Org 2} shares the encryption \texttt{secret} with \texttt{Org 3} using CYBEX-P KMS.

\subsubsection*{\textbf{Private Data Query}}\label{sss:privq}

\begin{figure}[ht]
	\includegraphics[height=.36in]{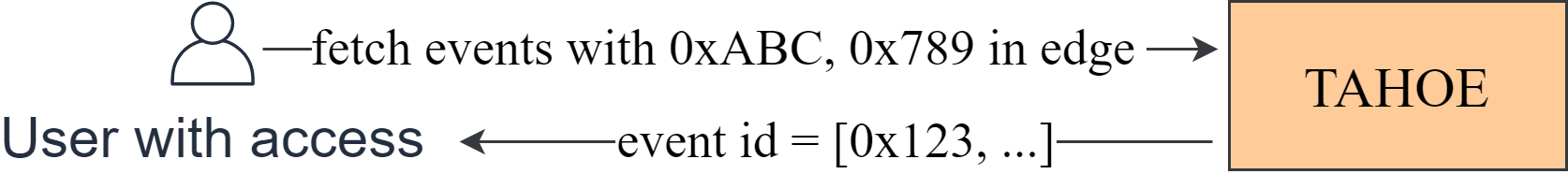} 
	\centering
	\caption{Private Data Query by Trusted Party.}
	\label{fig:privquetrusted}
\end{figure}

Now, \texttt{Org 3} wants to fetch all \texttt{events} with the IP \texttt{1.1.1.1}. She first generates the hash of \texttt{1.1.1.1} as \texttt{0xABC}. Then she encrypts the hash with \texttt{secret} to generate the ciphertext \texttt{0x789}. Finally, she queries the database for \texttt{events} that have \texttt{0xABC} or \texttt{0x789} in the \texttt{edge} array. The database will return the \texttt{event 0x123} along with public events which include \texttt{1.1.1.1}. Note that, this query still makes one pass over the database.

\subsubsection*{\textbf{Private Data Correlation}}

A powerful feature of CYBEX-P is intrinsic correlation of data as described in \ref{ss:fcorr}. What makes TAHOE even more powerful is that, the intrinsic correlation mechanism works on encrypted data as well.

As explained in subsubsection \ref{sss:privq}, an authorized user can query encrypted data without revealing its value. The query performs a graph traversal, returning a complete graph of all the related \texttt{attributes} and \texttt{events}. This graph contains all the intrinsic correlations described in \ref{ss:fcorr}.

\section{Implementation and Experimental Evaluation}\label{sec:implementation}


We have implemented CYBEX-P for experimental evaluation, with $5$ data sources, along with $4$ instances of MongoDB and $4$ servers to house the different modules. We have collected about $314$ billion events from several sources.

\subsection{Sources}

The sources we used for our demonstration are:

\begin{enumerate}
	\item Cyberthreat intelligence (CTI) from University of Nevada, Reno's MISP \cite{wagner2016misp} instance.
	\item SSH login attempts collected by four different instances of `cowrie' honeypot \cite{oosterhof2014cowrie}.
	\item Firewall log data from our honeypot system.
	\item Feed of phishing URLS from Phishtank \cite{phishtank}.
	\item Feed of phishing URLS from OpenPhish \cite{openphish}.
\end{enumerate}

\subsection{Complexity \& Scalability}\label{subsec:result}

\begin{figure}[h!]
	\centering
	\includegraphics[height=1.7in]{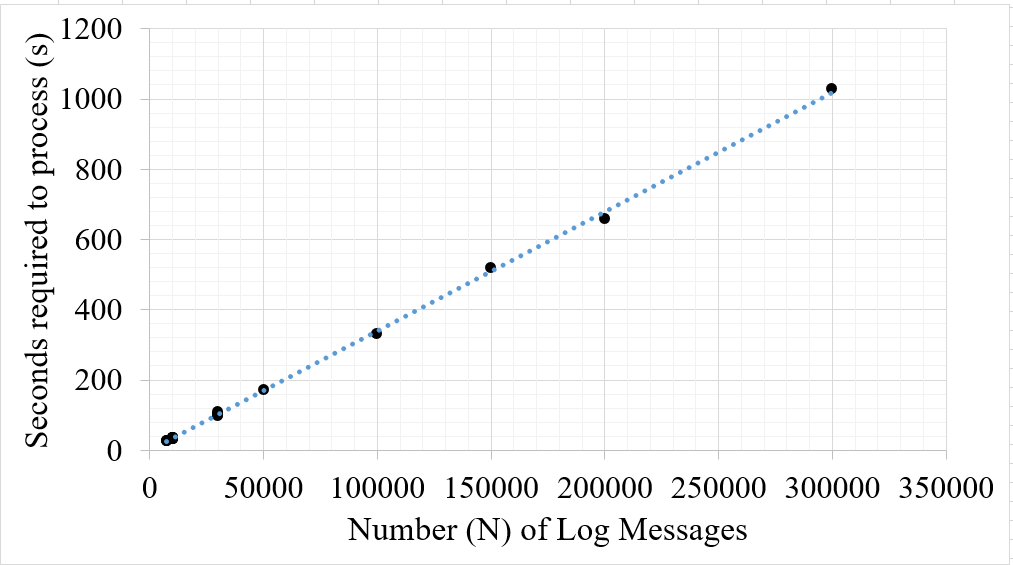}
	\caption{Evaluation of Complexity}
	\label{fig:eval}
\end{figure}

To test the scalability of CYBEX-P, we have fed $300000$ lines of {\it iptables} firewall log into it. Then we have recorded the time taken to process $N$ log messages. We have measured the time taken from input to storing in the archive DB as TAHOE events. A line of best fit is drawn among the data points. The result is shown in figure \ref{fig:eval}.

It can be inferred from the test that the overall complexity is linear. Furthermore, each log message can be processed independent of the other. This makes the system suitable for horizontal scaling using distributed computing paradigms.


\subsection{Data Compression by TAHOE}

\begin{figure}[!ht]
	\centering
	\includegraphics[width=3in]{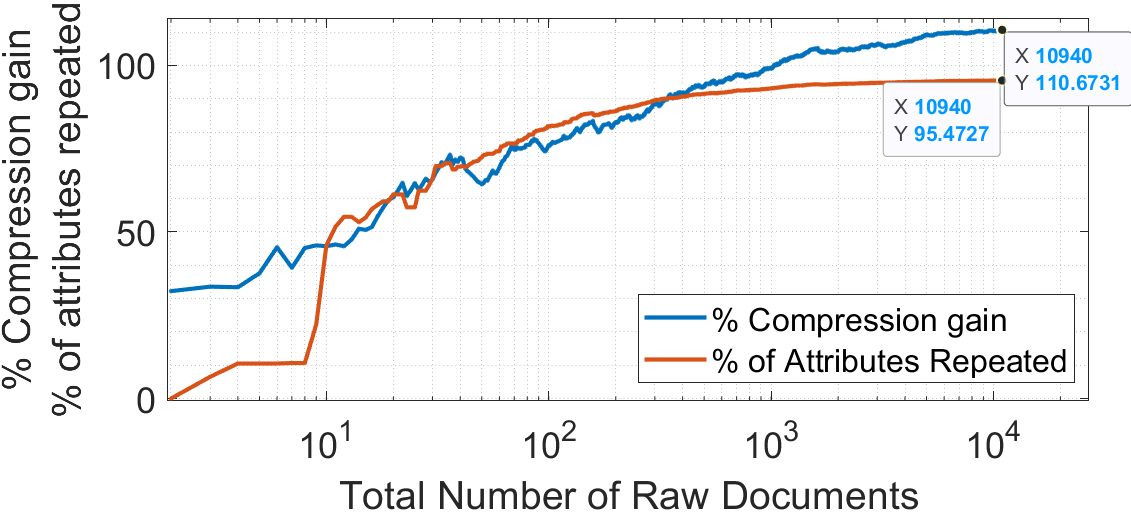}
	\caption{Data Compression in TAHOE}
	\label{fig:compression}
\end{figure}

As discussed in \ref{sss:dup}, TAHOE de-duplicates data, meaning there is only one instance of the IP \texttt{1.1.1.1} in our TAHOE database. Furthermore, \texttt{events} never store the actual value of an \texttt{attribute}, only a reference to it. The reference is the SHA256 hash of the attribute and only takes $32$ bytes of storage. So, if an \texttt{attribute} is repeated in another event, TAHOE takes only $32$ bytes of extra storage. As a result, TAHOE automatically achieves significant data compression as showed in Fig. \ref{fig:compression}.

As seen in Fig. \ref{fig:compression}, initially the compression gain is below $100\%$. However, as the percentage of repeated attributes grows so does the compression gain. Here, TAHOE achieves a compression gain of $10.7\%$ for only about $11$ thousand pieces of raw threat data, collected from our Cowrie honeypots.

\subsection{ThreatRank Verification by Case Study}\label{ss:respage}

In section \ref{sec:pagerank} we have formulated an algorithm called ThreatRank (TR) to detect malicious \texttt{events}. Here, we verify this algorithm. using the `Intrusion Kill Chain' \cite{hutchins2011intelligence} dataset from Lockheed Martin.

\subsubsection{\textbf{Intrusion Kill Chain and Correlation}}

Authors of \cite{hutchins2011intelligence} formulated the $7$ phases of an intrusion kill chain (also known as cyber kill chain) -- (1) Reconnaissance, (2) Weaponization, (3) Delivery, (4) Exploitation, (5) Installation, (6) Command and Control, (7) Actions. It is desirable to detect an attack in an early phase.

In their case study, their are $3$ related intrusion attempts. The first attempt delivered a malicious file via email. Although, at first the email looked benign, it was later flagged as malicious because it had a malicious attachment. Note that, by that time the attack has already passed phase $3$ undetected.

The next two intrusion attempts also delivered malicious files via emails. However, both these emails had similarities with the first email. As a result, the defenders could detect the attack even before analyzing the malicious files.

\begin{figure}[h!]
	\centering
	\includegraphics[height=1.2in]{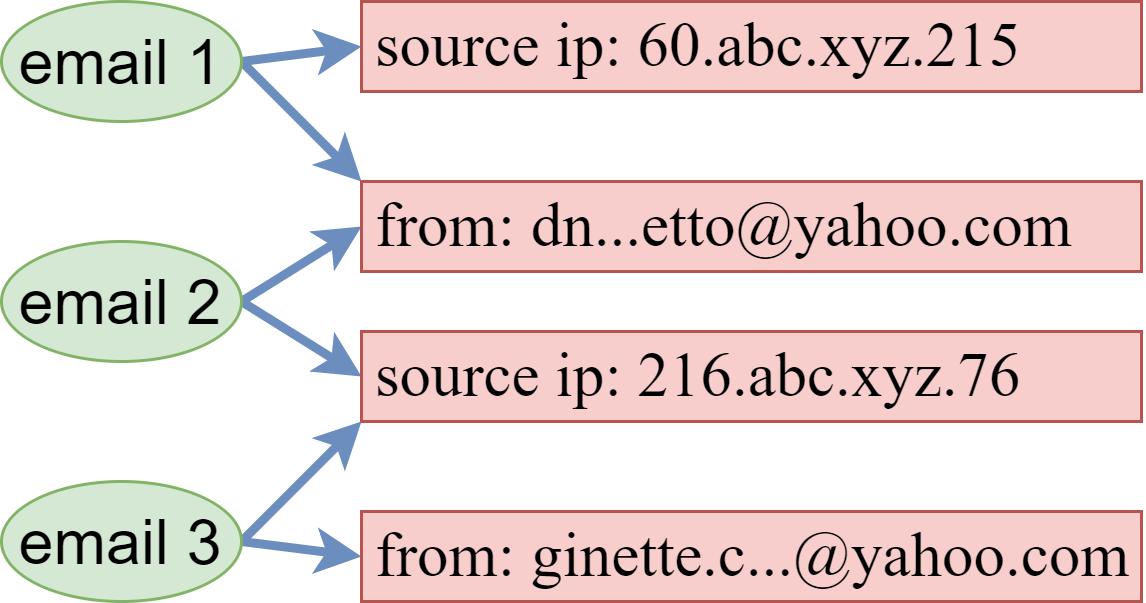} 
	\caption{Three emails from three separate intrusion attempts are intrinsically correlated in TAHOE because of common attributes.}
	\label{fig:lmdat}
\end{figure}

In other words, the defenders detected these two attempts in phase $3$, not later, because the emails are correlated. Fig. \ref{fig:lmdat} shows the common attributes in these emails as a TAHOE graph.

\subsubsection{\textbf{Automatic Intrinsic Correlation by TAHOE}}

While the correlations in Fig. \ref{fig:lmdat} are trivial, it is impossible for defenders to manually analyze all emails. TAHOE automates the process by intrinsically correlating the $3$ emails based on their common attributes as shown in Fig. \ref{fig:lmdat}.

However, correlating them is only half the battle. The correlation must be quantified before alerting the analysts. That is where ThreatRank steps in.

\subsubsection{\textbf{ThreatRank to Quantify Correlations}}

We assume that \texttt{email 1} has already been flagged as malicious in TAHOE database. So, we assign a fixed ThreatRank (TR) of $1$ to \texttt{email 1} and mark the two edges in Fig. \ref{fig:lmdat} as malicious. Then we simulate ThreatRank on the graph to get the results in Table \ref{tbl:TRres}. We have also added a benign email called \texttt{email 4} to the TAHOE database. \texttt{email 4} shares no common attribute with any of the \texttt{emails 1,2,3}.

\begin{table}[!ht]
	\renewcommand{\arraystretch}{1.3}
	\caption{ThreatRank of $4$ emails calculated on $3$ dates}
	\label{tbl:TRres}
	\centering
	\begin{tabular}{|c||c|c|c|}
		\hline
		Email & arrival$_2$ & arrival$_3$ & arrival$_1 + 365$ \\
		\hline
		\hline
		\texttt{email 1} & $-1$ & $-1$ & $-1$  \\
		\hline
		\texttt{email 2} & $-0.18$ & $-0.15$ & $-0.07$  \\
		\hline
		\texttt{email 3} & N/A & $-0.08$ & $-0.03$  \\
		\hline
		\texttt{email 4} & $0$ & $0$ & $0$ \\
		\hline
	\end{tabular}
\end{table}

Here, arrival$_1$ is the date of arrival of \texttt{email 1} and arrival$_1 + 365$ is one year later. Note that, \texttt{email 1} has fixed ThreatRank of $-1$ because it is already analyzed by an analyst. ThreatRank is calculated for unknown events only. Also, in the dataset \texttt{email 2} arrives $1$ day after \texttt{email 1} and \texttt{email 3} arrives $20$ days after \texttt{email 1}.

\texttt{email 2} has a TR of  $-0.18$ while \texttt{email 3} has a TR of $-0.08$ on respective arrival day. \texttt{email 3} has a lower TR because \texttt{email 2} is directly connected to \texttt{email 1}, whereas \texttt{email 3} is one hop away from \texttt{email 1}. Also, as expected their TR becomes almost half after a year. For all the simulations, TR of \texttt{email 4} remains $0$ because it shares no common attributes with the other emails.

\section{CYBEX-P Infrastructure as a Service (IaaS) for Phishing URL Detection}\label{sec:usecase}

As described in sections \ref{sec:intro} and \ref{sec:rw}, existing information sharing platforms have limited or zero support for data analysis. Our vision for CYBEX-P is to provide Infrastructure as a Service (IaaS) for all kinds of threat analysis. To augment this claim we have developed Phishly - a real-time phishing URL detector \cite{sadique2020automated} using CYBEX-P infrastructure. A phishing attack aims at stealing user information via deceptive websites.

\subsection{System Architecture}

\begin{figure}[ht!]
	\includegraphics[height=2.4in]{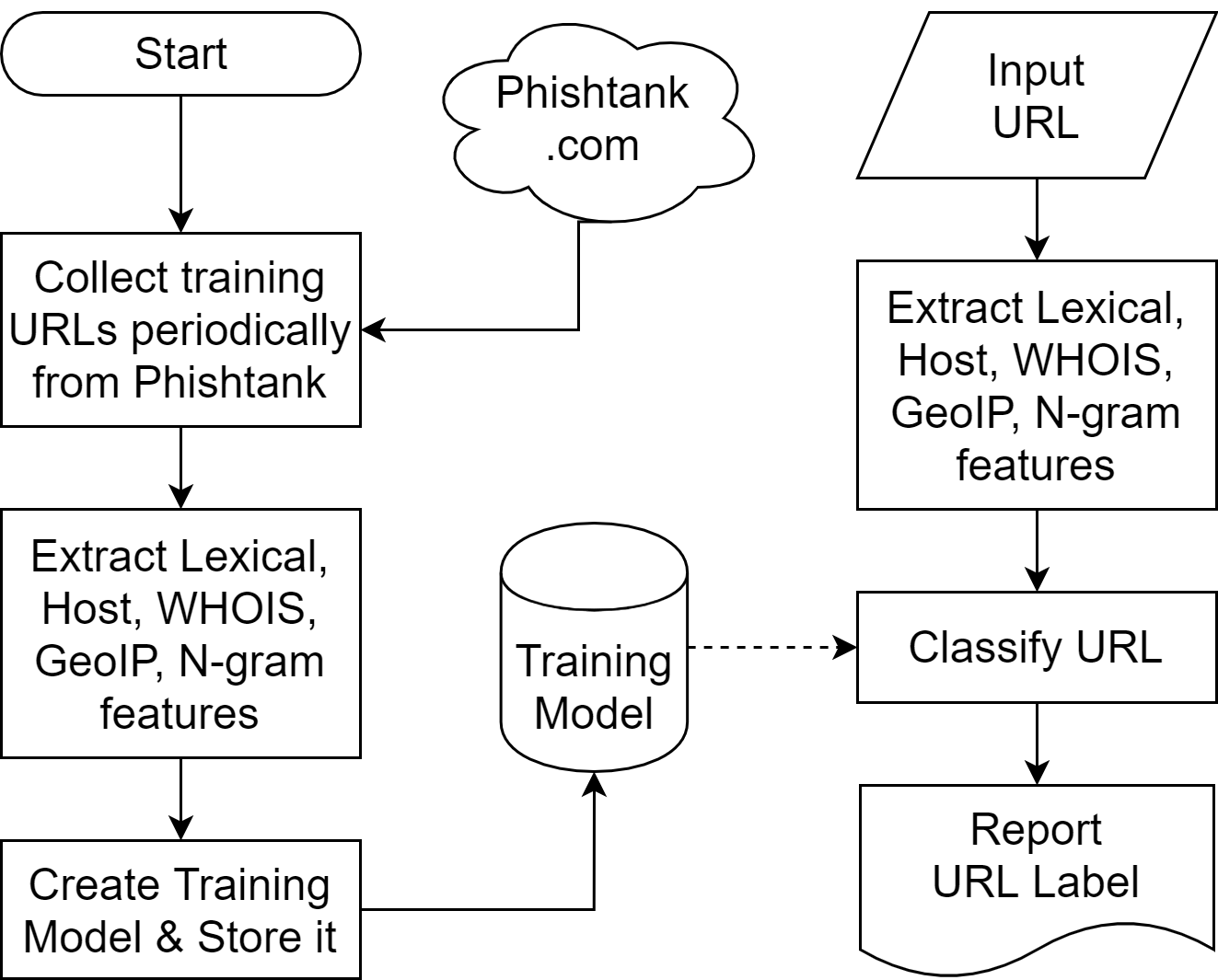}
	\centering
	\caption{System Architecture of the Phishing URL Detection system}
	\label{fig:phish}
\end{figure}

Fig. \ref{fig:phish} shows the system architecture of this system. As it uses CYBEX-P's infrastructure, it is explained using CYBEX-P components from Fig. \ref{fig:sysarch}.

Here, Phishtank.com stores raw URLs (\inlinefig{1} in Fig. \ref{fig:sysarch}). It provides an API (\inlinefig{2}) to share the URLs. Note that Phishtank's API is the connector from CYBEX-P's perspective. CYBEX-P collector (\inlinefig{3}) calls that API periodically to get new URLs and post them to the CYBEX-P API (\inlinefig{4}). CYBEX-P API puts them in the cache datalake (\inlinefig{5}). The archive cluster (\inlinefig{6.1}) parses the raw URLs into TAHOE events and stores them in the archive DB (\inlinefig{7}). The analytics cluster (\inlinefig{6.3}) enriches the URLs with $5$ sets of features. The feature sets are described in \ref{ss:urlfeat}. The analytics cluster also trains a classifier model.

On the other hand, a user can input an URL via the frontend (\inlinefig{10}, \inlinefig{9}) webapp. The analytics cluster classifies this URL, as benign or phishing, using the previously trained model. The report cluster (\inlinefig{6.2}) creates a report out of the classification label, and stores it in the report DB (\inlinefig{8}). The report is shown to the user via the frontend.

\subsection{Dataset}

We have collected about $36,000$ phishing URLs from Phishtank. We have also collected approximately $60,000$ benign URLs from Phishtank which were previously reported as suspicious but later analyzed to be benign. On the other hand, several previous works \cite{sahingoz2019machine, mamun2016detecting} used web crawling to generate benign URLs out of highly ranked websites only. This approach often results in a biased dataset.  Our benign URL dataset is better aligned with the real world.

\subsection{Features}\label{ss:urlfeat}

We extract five types of features from each URL: lexical, host, GeoIP, domain WHOIS and n-gram. Each feature set is described below:

\subsubsection{Lexical Features}

Lexical features are based on the URL string itself. Several examples of typical lexical features are number of characters in the URL, number of dots in the URL and number of symbols in the URL.

\subsubsection{Host Based Features}

Host based features are based on the server that hosts the webpage. The simplest such feature is the IP address that the URL resolves to.

\subsubsection{Domain WHOIS Based Features}

Any Regional Internet Registry (RIR) like ARIN or APNIC maintains it's own domain WHOIS database that contains information on the domain registrant. It contains the name of the registrant, the registration date, the expiration date etc. We query these data using RDAP \cite{rfc7842} protocol to get the domain WHOIS features.

\subsubsection{GeoIP Based Features}

GeoIP features are obtained from the IP address of the host. We used the GeoLite2 \cite{geolite2} IP geolocation databases from MAXMIND to collect various GeoIP features of the host IP address. GeoIP features include autonomous system number (ASN), country, city, latitutde, longitude etc.

\subsubsection{N-gram Features}

An n-gram is a continuous sequence of n characters from the URL. Phishing URLs often contain common brand names (like Microsoft, Paypal) to confuse the visitor. We catch those names using the URLs n-gram.

\subsection{Online Classification}\label{subsec:componl}

For this work, we have chosen the Second Order Perceptron (SOP) online classifier from the package LIBSOL \cite{wu2016libsol}. We have chosen an online classifier, because we collect new training URLs everyday. It is impossible to retrain our entire model with the total dataset because the dataset grows in an unbounded manner. The training model of the online classifier can be updated with only the new URLs without retraining it from the beginning. We have also compared the accuracy of SOP with batch random forest (RF) classifier from the package scikit-learn \cite{pedregosa2011scikit}.

\subsection{Result}

\begin{figure}[ht!]
	\includegraphics[height=1.8in]{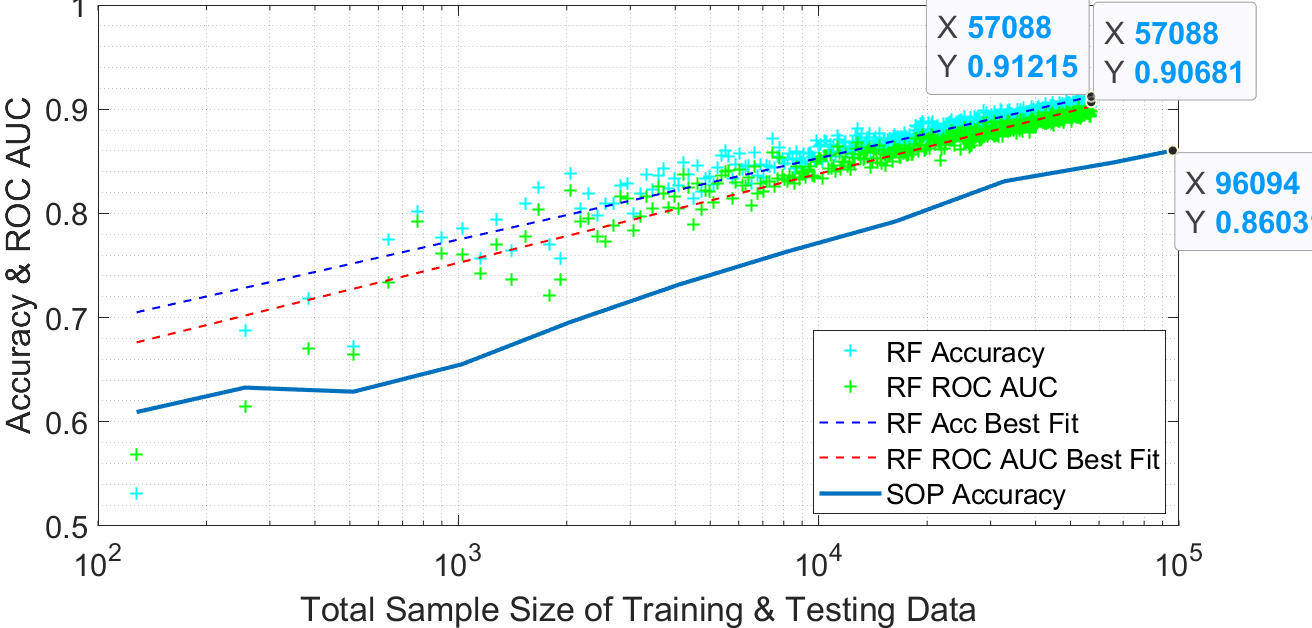}
	\centering
	\caption{Accuracy \& ROC AUC vs Sample Size}
	\label{fig:phishres}
\end{figure}

Fig. \ref{fig:phishres} shows the change in the accuracy and the `receiver operating characteristic - area under the curve' (ROC AUC) with growing sample size, for our system. As seen in the figure, for a sample size of $57$ thousand, batch RF achieves an accuracy of $0.91$ or $91\%$ with a ROC AUC of $90\%$. Online SOP, on the other hand, achieves an accuracy of $86\%$ for a sample size of about $96$ thousand. However, we still choose the online SOP because batch RF cannot deal with the unbounded growth of the URL dataset.

It can also be interpreted from the figure that the accuracy of our classifier will increase in future, as we get new data, because the curve has not become parallel to the x-axis yet.

\section{Conclusion \& Future Work}

In this paper, we have proposed CYBEX-P, as a completely automated cybersecurity information sharing (CIS) platform. We have also introduced TAHOE -- a graph based cyberthreat language (CTL), to overcome the limitations of existing CTLs. Moreover, we have introduced a universal Threat Data Query Language (TDQL) to facilitate sharing. Furthermore, we have formulated a novel algorithm called ThreatRank to detect malicious events. We have also tested the scalability and feasibility of CYBEX-P in a real world setup. Finally, we have showed how to use CYBEX-P infrastructure as a service (IaaS) with a phishing URL detection module. Our future goals for CYBEX-P are:

\begin{enumerate}
	\item Sequential analysis of related events (by timestamp).
	\item Real-time anomaly detection in a sequence of events.
	\item Provide cyber-insurance based on CYBEX-P.
	\item Detecting malicious actors who intentionally share bad data.
\end{enumerate}

\section{Acknowledgments}

\begin{acks}
This research is supported by the National Science Foundation (NSF), USA, Award \#1739032.
\end{acks}

\bibliographystyle{ACM-Reference-Format}
\bibliography{refs}

\end{document}